\documentclass[a4paper,11pt]{article}

\usepackage{jheppub} % for details on the use of the package, please see the JINST-author-manual
\usepackage{lineno}
\usepackage{float}

\title{Formation of holographic vortex in a rotating shell-shaped superfluid}

\allowdisplaybreaks[4]
\affiliation[a]{School of Physics and Astronomy, Beijing Normal University, Beijing 100875,
	China}
\affiliation[b]{Key Laboratory of Multiscale Spin Physics, Ministry of Education, Beijing Normal University, Beijing 100875, China}

\affiliation[c]{Department of Physics, Sichuan Normal University, Chengdu, Sichuan 610066, China}

\affiliation[d]{Zhiyuan School of Liberal Arts, Beijing Institute of Petrochemical Technology, Beijing 102617, China}

\affiliation[e]{School of Physical Sciences, University of Chinese Academy of Sciences, Beijing 100049, China}
\affiliation[f]{Institute of Theoretical Physics, Chinese Academy of Sciences, Beijing
	100190, China}

\author[a,b]{Chang-Xu Yan,}
\emailAdd{cxyan@mail.bnu.edu.cn}

\author[c]{Hao-Ran Chang,}
\emailAdd{hrchang@mail.ustc.edu.cn}

\author[d]{Meng Gao,}
\emailAdd{gaomeng@bipt.edu.cn}

\author[e,f]{Yu Tian,}
\emailAdd{ytian@ucas.ac.cn}

\author[a,b]{Hongbao Zhang}
\emailAdd{hongbaozhang@bnu.edu.cn}

\abstract{%Motivated by recent studies of homogeneous shell-shaped Bose–Einstein condensates (BECs), 
We investigate the holographic superfluid dynamics subjected to external rotation on a spherical geometry. 
Through a linear perturbation analysis, we identify several dynamically unstable phases in the phase diagram, each characterized by distinct unstable modes. Employing fully nonlinear numerical simulations, we further demonstrate that these unstable modes generically drive the system into vortex–antivortex configurations with definite winding numbers, determined by the symmetry of the corresponding unstable modes.}


\begin{document}
\maketitle

\section{Introduction}
 %as a hallmark of macroscopic quantum coherence, 
 Superfluidity, characterized by dissipationless flow and vanishing viscosity, has been a central topic in many-body physics since its discovery in liquid helium~\cite{allen1938flow,kapitza1938viscosity}. Among others, the advent of ultracold atomic gases has provided highly controllable platforms for exploring superfluid phenomena in Bose-Einstein condensates (BECs)~\cite{anderson1995observation,davis1995bose,raman1999evidence,madison2000vortex,abo2001observation,miller2007critical,henn2009observation,neely2010observation,ramanathan2011superflow,desbuquois2012superfluid,wright2013threshold,kwon2015periodic,kwon2016observation}. Moreover, BECs have greatly advanced the study of superfluidity while extending its influence to a broad range of disciplines, including quantum information science~\cite{greiner2002quantum,seaman2007atomtronics,bloch2012quantum}, strongly correlated and superconducting systems~\cite{greiner2003emergence,zwierlein2005vortices,giorgini2008theory}, and astrophysical phenomena~\cite{steinhauer2016observation,pines1985superfluidity,almeida2023analogue}.
Their remarkable tunability has enabled the realization and investigation of superfluid systems in a wide range of geometries and dimensionalities, including disk-shaped, cigar-shaped, toroidal, and spherical-shell configurations. While most experimental and theoretical studies have focused on planar geometries, shell-shaped systems have attracted growing interest in recent years. This interest has been driven by the development of several microgravity platforms for cooling and trapping ultracold atomic gases in microgravity environments~\cite{elliott2018nasa,aveline2020observation,van2010bose,muntinga2013interferometry,vogt2020evaporative,becker2018space,condon2019all}, where the realization of closed shell-shaped BECs was anticipated~\cite{lundblad2019shell} and has subsequently been achieved aboard the Cold Atom Lab (CAL) on the International Space Station~\cite{carollo2022observation}. Complementary Earth-based experiments have also realized shell-shaped condensates in immiscible dual-species BECs~\cite{jia2022expansion}. These experimental advances have stimulated extensive theoretical investigations of shell-shaped BECs, including their superfluid properties~\cite{tononi2019bose,tononi2020quantum,kanai2021true,moller2020bose,tononi2022topological,tononi2022scattering,ciardi2024supersolid}, vortex dynamics~\cite{padavic2020vortex,bereta2021superfluid,white2024triangular}, and collective excitations~\cite{lannert2007dynamics,padavic2017physics,sun2018static,rhyno2021thermodynamics,li2023equatorial,saito2023rossby,boegel2023controlled,lundblad2023perspective}. However, these theoretical approaches suffer from several important limitations. For example, the widely used Gross-Pitaevskii (GP) equation can only describe the ground-state dynamics of BECs, while finite-temperature effects induced by the surrounding thermal bath are usually incorporated phenomenologically through dissipative terms added by hand~\cite{hohenberg1977theory}. On the other hand, conventional statistical approaches to finite-temperature BECs are restricted to equilibrium thermodynamic properties. Furthermore, although superfluid hydrodynamics provides an effective description of near-equilibrium dynamics, it is generally inadequate for far-from-equilibrium processes, where topological defects are ubiquitous and play a crucial role. Compared to GP approach, holographic duality, as a first-principles framework for us to address the finite temperature superfluid dynamics by mapping it into a gravitational system in an Anti-de Sitter (AdS) black hole with one extra dimension \cite{PhysRevLett.101.031601,hartnoll2008holographic}, has a universal applicability to more generic circumstances \cite{liu2019holographic}. In fact, since its advances, the holographic model of superfluids has been successfully applied to a wide range of problems in superfluid dynamics, encompassing characteristic length scales and universal non-equilibrium behaviors associated with topological defects such as dark solitons and vortices~\cite{keranen2010inhomogeneoussoliton,keranen2010inhomogeneous,domenech2010emergent,keranen2011solitons,montull2012magnetic,xia2019vortex,lan2019attractive,guo2020dynamical,li2020generation,wittmer2021vortex,ewerz2021dynamics,yan2023holographic,lan2023splitting,lan2023heating,yang2023motion,su2023giant}, as well as quantum turbulence~\cite{chesler2013holographic,du2015holographic,ewerz2015non,lan2016towards,yang2024mechanism,zeng2025dissipation}. Recently, the holographic analysis of the critical temperature in homogeneous shell-shaped superfluids~\cite{gao2025holographic} was found to be consistent with the previous results for superfluids obtained through conventional statistical approaches~\cite{tononi2019bose,tononi2020quantum,rhyno2021thermodynamics}. %Furthermore, this consistency paves the way for extending the study to more general cases involving external rotation, enabling the exploration of both the critical temperature and the associated non-equilibrium dynamics.
With this in mind, we intend to apply such a holographic model of superfluid to explore the shell-shaped superfluid dynamics under external rotation.

This paper is organized as follows. In Section~\ref{model}, we briefly review the holographic superfluid model. Section~\ref{configuration} is devoted to the construction of static solutions for inhomogeneous shell-shaped superfluids under external rotation. We also analyze the dependence of the critical temperature on the angular velocity at fixed particle number, followed by a linear instability analysis at various temperatures and angular velocities. Several fully nonlinear dynamical evolutions triggered by unstable modes are investigated in Section~\ref{non_linear}. Finally, we conclude in Section~\ref{conclusion}.

\section{Holographic setup}\label{model}
A simplest holographic superfluid model is described by the Abelian-Higgs model coupled to Einstein's gravity in the asymptotically AdS spacetime, the corresponding action is given by \cite{PhysRevD.78.065034,PhysRevLett.101.031601,PhysRevD.79.066002}
 \begin{align}
 	S=\frac{1}{16\pi G}\int_\mathcal{M}d^4x\sqrt{-g}\left[\left(R+\frac{6}{L^2}\right)-\frac{1}{q^2}\left(|D_\mu \Psi|^2+m^2|\Psi|^2+\frac{1}{4}F^{2}_{\mu\nu}\right)\right],
 \end{align}
 where G is the Newtonian gravitational constant, R is the scalar curvature of spacetime, L is AdS radius. $D_\mu=\nabla_\mu-i A_\mu$, with $\nabla_{\mu}$ the covariant derivative operator. $\Psi$ is a complex scalar field coupled to the gauge field $A_\mu$, with mass $m$ and charge $q$. In what follows, we almost focus on the probe limit, namely $q\to\infty$, so that the backreaction of the matter fields to the background metric can be ignored. To address the superfluid dynamiccs on the shell, the spherically symmetric Schwarzschild-$\text{AdS}_4$ spacetime is considered:
 
 \begin{align} \label{background}
 	ds^2=\frac{L^2}{z^2}\left[-f(z)dt^2+\frac{dz^2}{f(z)}+L^2(d\theta^2+\mathrm{sin}^2(\theta)d\phi^2)\right],
 \end{align} 
 where $f(z)=1+\frac{z^2}{L^2}-\left(\frac{z}{z_h}\right)^3\left(1+\frac{z_h^2}{L^2}\right)$ with $ z=z_h$ the horizon location. Hawking temperature can be easily extracted as
 \begin{align}
 	T_H=\frac{|f^{\prime}(z_h)|}{4\pi}=\frac{3+z_h^2/L^2}{4 \pi z_h},
 \end{align} 
 which can be alternatively interpreted as the temperature of dual boundary system living on the sphere with radiu $L$. For convenience, we adopt the natural units $(c=1,G=1,\hbar=1)$. Apparently,
 the temperature reach  its minimal value $T_{min}=\frac{\sqrt{3}}{2\pi L}$ at $z_h=\sqrt{3} L$. Bellow the $T_{min}$, there is no black hole solution, on the contrary, when $T\ge T_{min}$, there exist two black hole solutions, commonly referred to as the large black hole $(z_h<\sqrt{3} L)$ and the small black hole $(z_h>\sqrt{3} L)$, respectively. However, the thermodynamically stable state is given by the large black hole. So in what follows, we would like to take the large one as our background spacetime. Moreover, without loss of generality, we shall work in the unit in which $L=1$.

The dynamics of the bulk matter fields are controlled by the following equations of motion
\begin{align}
	&D_{\mu}D^{\mu}\Psi-m^2\Psi=0,\notag\\
    &\nabla_\mu F^{\mu\nu}=J^{\nu},
	\label{EOM}
\end{align}
with $J^{\nu}=i\left[\Psi^{*} D^\nu\Psi-\Psi \left(D^{\nu} \Psi\right)^{*}\right]$. 
To be more specific, we set $m^2= -2$ and adopt the axial gauge $A_z = 0$. Accordingly, the bulk matter fields exhibit the following asymptotic behaviors near the AdS boundary $z=0$
\begin{align}
	&\Psi= z \left(\Psi_{-}+\Psi_{+} z+\cdots\right),\notag\\
    &A_\nu= a_\nu+j_\nu z+\cdots.
\end{align} 
Here $j_\nu=(-\rho,j_i)$ is interpreted as the $U(1)$ conserved current for the boundary system sourced by $a_{\nu}=(\mu,a_i)$ with $\rho$ and $\mu$ corresponding to the number density and chemical potential, respectively. 
In addition, $\Psi_{+}$ is interpreted as the expectation value of the condensate operator $O$ sourced by $\Psi_-$.  With $\Psi_{-}=0$, the condensate  $\langle O\rangle$ becomes nonzero, signaling
the onset of $U(1)$ spontaneous symmetry breaking to the
superfluid phase from the normal fluid phase.

 \section{Linear instability of a rotating shell-shaped superfluid}\label{configuration}	
 \subsection{Rotating shell-shaped superfluid}
  To construct the rotating shell-shaped superfluid, we would like to make the following ansatz for the bulk matter fields
\begin{align}
	\Psi=z \psi(z,\theta), \quad A_\nu=A_\nu(z,\theta)
	\label{anstz}
\end{align} 
where $A_\theta=0$, whereby the resulting equations of motion can be written explicitly as
	\begin{align}
		&\partial_z(f\partial_{z}\psi) + \frac{2-2f+z\partial_zf}{z^2}\psi+\frac{\partial_{\theta}(\mathrm{sin}\theta\partial_{\theta}\psi)}{\mathrm{sin}\theta}+\left(\frac{A_{t}^2}{f}-\frac{A_{\phi}^2}{\mathrm{sin}^2\theta}\right)\psi=0,\\
		&f\partial_{z}^{2}A_{\phi}+\partial_{z}f \partial_{z}A_{\phi}-2A_{\phi}\psi^2+\partial_{\theta}^{2}A_\phi-\mathrm{cot}\theta\partial_{\theta}A_{\phi}=0,\\
		&f\partial_{z}^{2}A_t-2A_{t}\psi^2+\partial_{\theta}^{2}A_{t}+\mathrm{cot} \theta\partial_{\theta}A_{t}=0.
	\end{align}
    
\begin{figure}
\centering
		\includegraphics[width=0.5\linewidth]{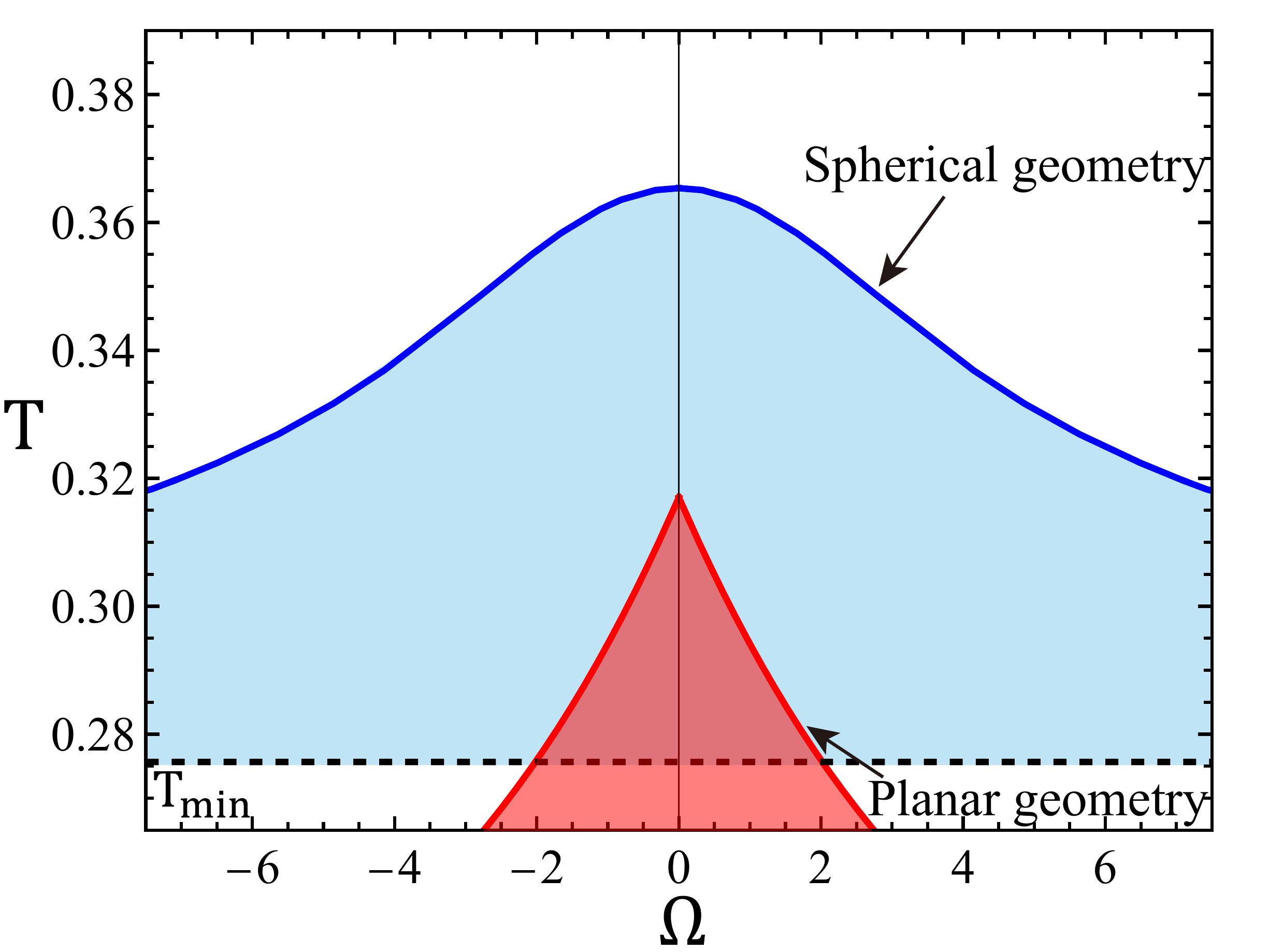}
		\caption{Comparison between the critical lines for the onset of superfluidity on the sphere (blue) and on the plane (red), shown in the $T$–$\Omega$ phase diagram at $\bar\rho=22.5/\pi$.}
		\label{diff}
	\end{figure}
Because we work in the canonical ensemble, the average number density on the unit sphere, $\bar{\rho}$, is held fixed and is defined by
\begin{align}
\bar\rho=\frac{1}{4\pi}\int_{S^2} dS~\rho
=-\frac{1}{4\pi}\int_{S^2} dS~\partial_z A_t|_{z=0}.
\end{align}
But, for the chemical potential $\mu$, we only require $A_t|_{z=0}$ to be homogeneous on the sphere, without fixing its value a priori. Instead, it is determined self-consistently by the equations of motion. In addition, we also impose the following boundary conditions on the AdS boundary $z=0$,
	\begin{align}
		\psi|_{z=0}=0, A_{\phi}|_{z=0}=-\Omega~\mathrm{sin}^2\theta,
	\end{align}
where $\Omega$ is related to the prescribed external angular velocity $\omega$ of the rigid rotation of the shell
as $\Omega=\mu\omega$ and the resulting linear velocity is given
by $\upsilon=\frac{\Omega}{\mu}\mathrm{sin}\theta$. In addition, the regular boundary conditions are imposed on the horizon. On the other hand, at $\theta=0$ and $\theta=\pi$ from the AdS boundary all the way to the black hole horizon within the bulk, we require
	\begin{align}
		\partial_\theta\psi=0,\quad A_\phi=0, \quad \partial_{\theta}A_{t}=0.
	\end{align}
	Using the pseudo-spectral method with 26 Chebyshev modes in the $z$ direction and 30 Chebyshev modes in the $\theta$ direction, we compute the static configuration. The resulting solution exhibits equatorial symmetry, and changing the sign of $\Omega$ leaves the solution invariant except for an overall sign flip in $A_\phi$

As illustrated in figure~\ref{diff}, at a fixed average number density $\bar\rho = 22.5/\pi$, there exists a critical temperature (blue curve) that marks the onset of the shell-shaped superfluid configuration. As the magnitude of the angular velocity increases, the critical temperature decreases, as expected, since the rotation-induced superflow tends to spoil superfluidity according to the Landau criterion \cite{amado2014holographic,gouteraux2023critical,g8sg-prdf}. Furthermore, we also show the critical temperature (red curve) for rotating superfluid in planar geometry. We not only revisit previous observations, based on statistical-mechanical approaches \cite{tononi2019bose,tononi2020quantum,rhyno2021thermodynamics} and holography \cite{gao2025holographic}, that the critical temperature of a non-rotating superfluid is enhanced on the sphere compared to its planar counterpart, but also extend this conclusion to the general rotating case. In particular, we find that the enhancement becomes more pronounced at larger angular velocities.

\begin{figure}
		\includegraphics[width=\linewidth]{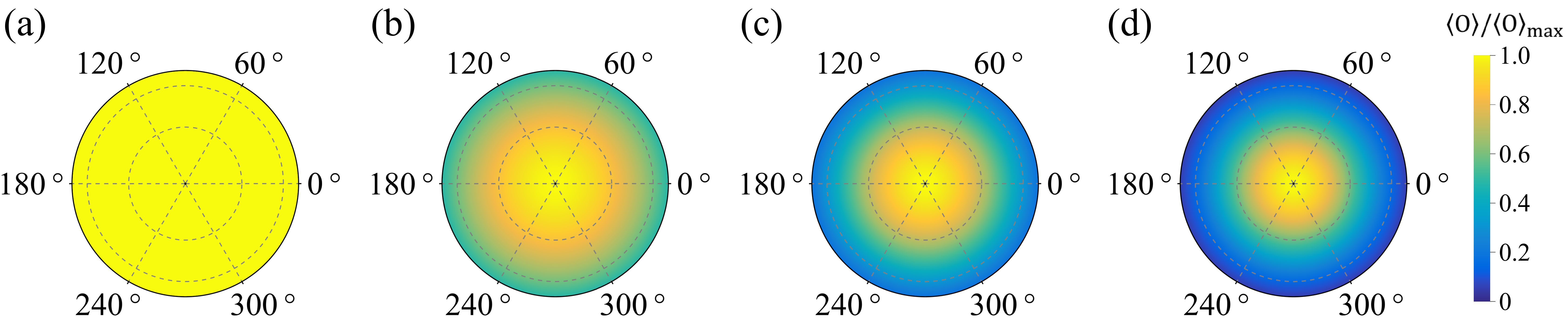}
		\caption{Superfluid on the northern hemisphere at $\bar\rho = 22.5/\pi$ and $T = 0.32$. Panels (a)–(d) show the order-parameter profiles for different angular velocities with $\Omega = 0, 3.8, 5.2$, and $6.6$, individually.}
		\label{static}
	\end{figure}

As shown in figure \ref{static}, we present normalized condensate amplitude $\langle O \rangle / \langle O \rangle_{\mathrm{max}}$ on the northern hemisphere at $\bar\rho = 22.5/\pi$ and $T = 0.32$ where panels (a)–(d) correspond to $|\Omega| = 0$, $3.8$, $5.2$, and $6.6$, individually. The distribution on the southern hemisphere is identical to that on the northern, thus we only need to present the superfluid configuration on the northern hemisphere without loss of generality. We observe that $\langle O \rangle / \langle O \rangle_{\mathrm{max}}$ on the equator decreases as the magnitude of the angular velocity increases, and this is the expected result when the rotating superfluid consist of no vortex. When the angular velocity becomes sufficiently large, the system develops a dynamical instability under perturbations and undergoes a phase transition, during which vortices nucleate at the locations where the condensate amplitude reaches its minimum \cite{xia2019vortex,li2020generation}.  In the following, we identify the critical $\Omega s$ at which the dynamical instability happen and investigate the nonlinear dynamics of the rotating shell-shaped superfluid.

\subsection{Quasi-normal modes}\label{secQNM}
In this section, we investigate the linear instability of rotating shell-shaped superfluid, which can be achieved by calculating its quasi-normal modes (QNMs). As such, it turns out to be a great convenience to work in the in-going Eddington-Finkelstein coordinates, where the ingoing boundary conditions at the horizon amounts to the regularity over there. Accordingly, the background metric takes the following form
\begin{align}
	ds^2=\frac{1}{z^2}\left(-f(z)dt^2-2dtdz+d\theta^2+\mathrm{sin}^2\theta d\phi^2\right).
\end{align}  

In order to preserve the axial gauge $A_{z}=0$ in the above new coordinates for the static configuration  previously obtained , we are required to additionally perform a proper $U(1)$ gauge transformation as follows
\begin{align}
	\psi(z,\theta)\to e^{i\lambda}\psi(z,\theta), \quad \partial_z \lambda=-\frac{A_t}{f}.
\end{align}

Since the resulting static configuration possess both the time translation and rotation symmetries, the linear perturbations of the matter fields can be constructed as
\begin{align}\label{perturbation}
	&\delta\psi=q_1(z,\theta)e^{-i \omega t+i p\phi}+ q_2^{*}(z,\theta)e^{i \omega^{*} t-i p\phi},\\ 
	&\delta A_{\mu}=a_\mu(z,\theta) e^{-i \omega t+i p \phi}+a_\mu^{*}(z,\theta) e^{i \omega^*t-i p \phi},
\end{align} 
where the coefficients $q_1$, $q_2$ and $a_{\mu}$ are all complex functions with the Dirichlet boundary conditions imposed on the AdS boundary $z=0$. By applying the above perturbations on top of the aforementioned static configuration and substituting them into eq.~(\ref{EOM}), we obtain the linearized perturbation equations for coefficients as
\begin{align}
		&\left(2i(\omega+A_t)\partial_z +i\partial_{z}A_{t}+i \partial_z(f\partial_z)+(\partial_{\theta}-i A_{\theta})^2+\frac{\partial_{\theta}-iA_{\theta}}{\mathrm{tan}\theta}-\frac{(p-A_{\phi})^2}{\mathrm{sin}^2\theta}-z\frac{1+z_h^2}{z_h^3}\right) q_1
		\notag\\
		&+(i\psi\partial_z+2 i\partial_{z}\psi)a_t-\left(i\psi\partial_{\theta}+2i\partial_{\theta}\psi+2A_{\theta}\psi+i\frac{\psi}{\mathrm{tan}\theta}\right)a_{\theta}+\frac{\psi(p-2A_{\phi})}{\mathrm{sin}^2\theta}a_{\phi}=0,\\ \notag\\
		%%%%%%%%%%%%%%%%%%%%%%%%%%%%%%%%%%%%%%%%%%%%%%
		&\left(2i(\omega-A_t)\partial_z-i\partial_{z}A_{t}-i \partial_z(f\partial_z)+(\partial_{\theta}+i A_{\theta})^2+\frac{\partial_{\theta}+i A_{\theta}}{\mathrm{tan}\theta}-\frac{(p+A_{\phi})^2}{\mathrm{sin}^2\theta}-z\frac{1+z_h^2}{z_h^3}\right)q_2 \notag\\
		&-(i\psi^{*}\partial_z+2 i\partial_{z}\psi^{*})a_t+\left(i\psi^{*}\partial_{\theta}+2i\partial_{\theta}\psi^{*}-2A_{\theta}\psi^{*}+i\frac{\psi^{*}}{\mathrm{tan}\theta}\right)a_{\theta}-\frac{\psi^{*}(p+2A_{\phi})}{\mathrm{sin}^2\theta}a_{\phi}=0,\\
		\notag\\
		%%%%%%%%%%%%%%%%%%%%%%%%%%%%%%%%%%%%%%%%%%%%%%
		&\frac{\partial_{z}a_{\theta}}{\mathrm{tan}\theta}+\partial_{z}\partial_{\theta}a_{\theta}-\partial_{z}^{2}a_{t}+i \frac{p}{\mathrm{tan}^{2}\theta}\partial_{z}a_{\phi}+i(\psi^{*}\partial_{z}q_{1}-q_1\partial_{z}\psi^{*})+i(q_2\partial_{z}\psi-\psi\partial_{z}q_{2})=0,\\ \notag\\
		%%%%%%%%%%%%%%%%%%%%%%%%%%%%%%%%%%%%%%%%%%%%%%%
		&\left(i f (\psi^{*}\partial_ {z} - \partial_ {z}\psi^{*}) - (\omega + 2 A_ {t})\psi^{*}\right) q_ {1} + \left(if (\partial_ {z}\psi - \psi \partial_ {z}) + (\omega - 2 A_ {t})\psi\right) q_ 2+\frac{p}{\mathrm{sin}^{2}\theta}(if\partial_{z}-\omega a_{\phi}) \notag \\
		&+\left(i\omega\partial_{z}+\partial_{\theta}^2+\frac{\partial_{\theta}}{\mathrm{tan}\theta}-\frac{p^2}{\mathrm{sin}^{2}\theta}-2\psi\psi^{*}\right)a_{t}+(f\partial_{z}+i\omega)\left(\partial_{\theta}+\frac{1}{\mathrm{tan}\theta}\right)a_{\theta}=0,\\ \notag\\
		%%%%%%%%%%%%%%%%%%%%%%%%%%%%%%%%%%%%%%%%%%%%%%%
		&\left(-2 i\omega\partial_ {z} - f\partial_ {z}^2 - 
		f^{\prime}\partial_ {z} + \frac{p^2}{\mathrm {sin}^2 \theta} + 
		2\psi\psi^{*}\right) a_ {\theta}+i \frac{p}{\mathrm {sin}^2 \theta} \partial_ {\theta} a_ {\phi}-\partial_ {z}\partial_ {\theta} a_ {t}\notag\\ 
		&+(i \psi^{*}\partial_ {\theta} - i\partial_ {\theta}\psi^{*} + 
		2\psi^{*} A_ {\theta}) q_ {1} +(-i\psi\partial_ {\theta} + 
		i\partial_ {\theta}\psi + 
		2 A_ {\theta}\psi) q_ 2=0,\\ \notag\\
		%%%%%%%%%%%%%%%%%%%%%%%%%%%%%%%%%%%%%%%%%%%
		&\left(2i\omega\partial_ {z} + \partial_ {z} (f\partial_ {z})+\partial_{\theta}^2-\frac{\partial_ {\theta}}{\mathrm{tan}\theta}-\psi\psi^{*}\right)a_ {\phi}- 
		ip\left(\partial_ {\theta}-\frac{1}{\mathrm{tan}\theta}\right)a_{\theta}+ip\partial_{z}a_{t}\notag\\
		&+(p-2A_{\phi})\psi^{*} q_1-(p+2A_{\phi})\psi q_{2} = 0.
	\end{align}

    The fourth equation evaluated on the AdS boundary and 
	\begin{align}
		q_1|_{z=0}=0,\quad q_2|_{z=0}=0,\quad a_t|_{z=0}=0,\quad a_\theta|_{z=0}=0,\quad a_\phi|_{z=0}=0
	\end{align}
    serve as the boundary conditions on the AdS boundary. In addition, we only require the regular boundary conditions on the horizon. At $\theta=0$ and $\theta=\pi$ within the bulk, we impose 
	\begin{align}
		\partial_{\theta}q_1=0,\quad \partial_{\theta}q_2=0,\quad \partial_{\theta}a_{t}=0,\quad a_{\theta}=0, \quad
		\partial_{\theta}a_{\phi}=0.
	\end{align}
	
The QNMs can be obtained by solving the generalized eigenvalue problem. A mode with the corresponding frequency $(\omega )$ having a
positive imaginary part indicates the linear instability of
the system under the perturbation of such a mode. 

\begin{figure}
\centering
		\includegraphics[width=0.8\linewidth]{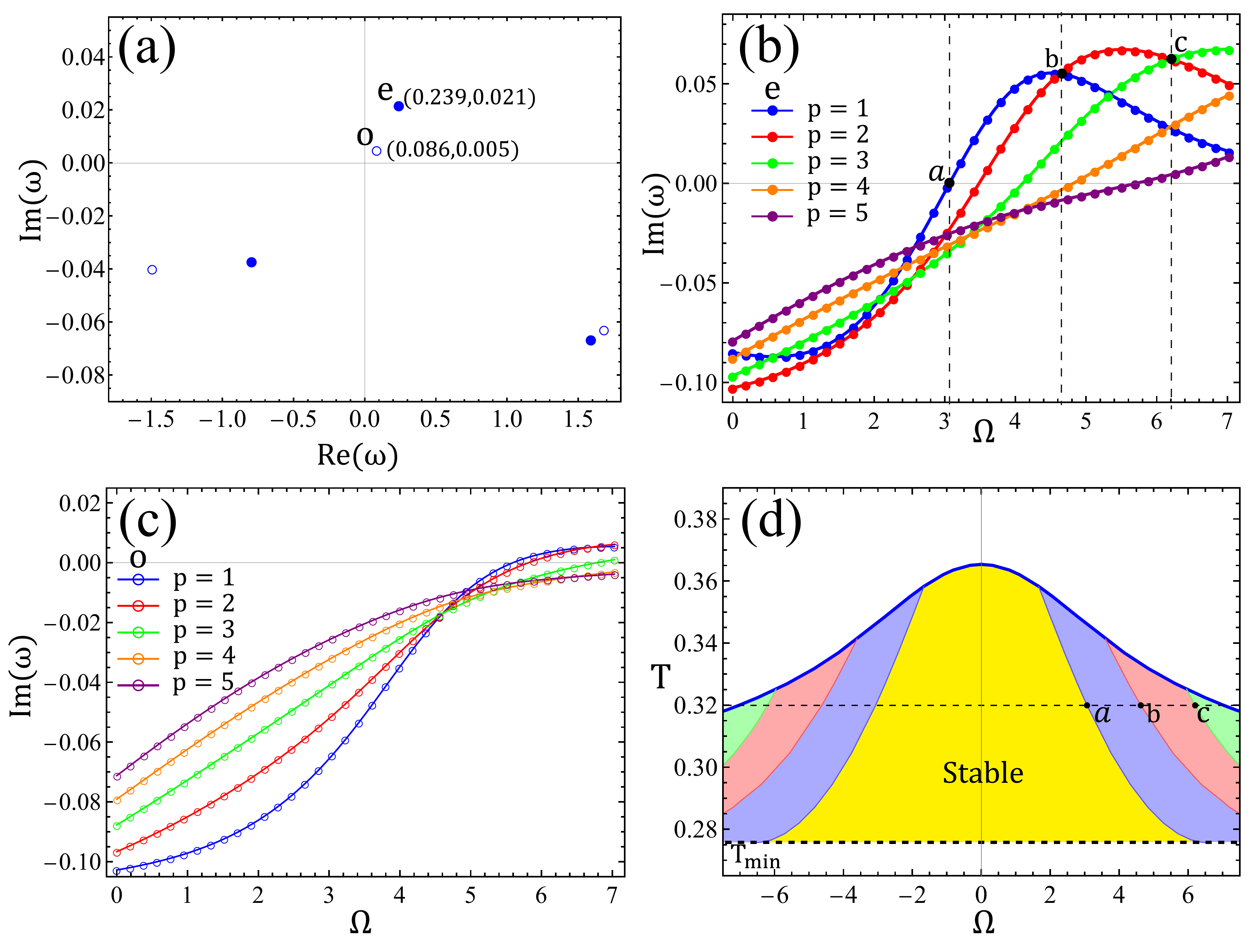}
		\caption{(a) The spectrum of QNMs of the superfluid at $\bar\rho=22.5/\pi$, $T=0.32$ and $\Omega=6.6$ for $p=1$, where the even and odd modes are marked in desk and in circle, respectively. (b)-(c) The $\Omega$-dependence of $\mathrm{Im}(\omega)$ for the even and odd QNMs in the channel $p=1,2,3,4,5$, at $\bar\rho=22.5/\pi$ and $T=0.32$. (d) Dynamical phase diagram in the $T-\Omega$ plane for the  superfluid on the sphere at $\bar\rho=22.5/\pi$, where the dynamical processes for the blue, red and green regions are controlled by the dominant unstable even modes with $p=1$, $2$, and $3$, individually.}
		\label{QNM_N90}
	\end{figure} 

As an illustration, we still stick to $\bar\rho=22.5/\pi$ and present the spectrum of QNMs for $p=1$ in figure~\ref{QNM_N90}(a) on top of the superfluid configuration at $T = 0.32$ and $\Omega=6.6$, presented in figure~\ref{static}(d). As one can
see, the resulting QNMs can be classified into even and odd modes, which satisfy
\begin{align}
&\delta\psi(\theta)=\delta\psi(\pi-\theta),\quad\delta A_t(\theta)=\delta A_t(\pi-\theta),\notag\\
&\delta A_\phi(\theta)=\delta A_\phi(\pi-\theta),\quad\delta A_\theta(\theta)=-\delta A_\theta(\pi-\theta),
\end{align}
and 
\begin{align}\label{symmetry_for_odd_mode}
&\delta\psi(\theta)=-\delta\psi(\pi-\theta),\quad\delta A_t(\theta)=-\delta A_t(\pi-\theta),\notag\\
&\delta A_\phi(\theta)=-\delta A_\phi(\pi-\theta),\quad\delta A_\theta(\theta)=\delta A_\theta(\pi-\theta),
\end{align}
respectively.

By fixing the temperature at $T = 0.32$, we
can further draw the dependence of the imaginary part of $\omega$ on $\Omega$ for both even and odd modes, as shown in figure~\ref{QNM_N90}(b) and (c), respectively. Both even and odd modes can provide dynamically unstable channels, however, the even modes are dominant over the odd ones. Accordingly, we classify the unstable phases in the phase diagram based on the dominant unstable even  modes with different values of $p$. We can identify three critical $\Omega s$, denoted by $\Omega_{a}$, $\Omega_{b}$, and $\Omega_{c}$ in figure~\ref{QNM_N90}(b). The first, $\Omega_{a}=3.067$, marks a critical point, beyond which dynamical unstable happens. The other two critical values, $\Omega_b=4.627$ and $\Omega_c=6.205$, correspond to the points at which the even modes in the $p=2$ and $p=3$ channels, respectively, overtake all other unstable modes to become the dominant instability. By determining the temperature dependence of these three critical angular velocities, we further divide the dynamical phase diagram of the superfluid into seven regions, as shown in figure~\ref{QNM_N90}(d).

It should be noted that QNMs corresponding to superfluid with negative values of $\Omega$, denoted by $\omega_n$, differ slightly from their counterparts $\omega_p$ for positive $\Omega$ and satisfy the relation $\omega_p = - \omega_n^{*}$. Consequently, the two spectra have identical imaginary parts, giving rise to a phase diagram symmetric about $\Omega=0$. However, their real parts have opposite signs, leading to subtly different dynamical evolutions, as will be demonstrated below.

\section{Dynamical process of phase transition}\label{non_linear}
In the previous section, we established the linear instability of the rotating shell-shaped superfluid. We now perform fully nonlinear numerical simulations of the bulk dynamics, starting from the static bulk scalar configuration perturbed by the dominant unstable even modes in the $p=1$, $2$, and $3$ channels, as well as the unstable odd modes in the $p=1$ and $p=2$ channels. The resulting fully nonlinear equations of motion for the bulk matter fields are given explicitly as follows
\begin{align}
	2\partial_{t}\partial_{z}\psi&=\partial_{z}(f\partial_{z}\psi+i A_{t}\psi)+ i A_{t}\partial_{z}\psi+\frac{\partial_{\theta}-iA_{\theta}}{\mathrm{tan}\theta}\psi+(\partial_{\theta}-i A_{\theta})^2\psi+\frac{(\partial_{\phi}-i A_{\phi})^2}{\mathrm{sin}^{2}\theta}\psi-z\frac{1+z_h^2}{z_h^3}\psi\label{Phi}\\
%%%%%%%%%%%%%%%%%%%%%%%%%%%%%%%%%%%%%%%%%%%%%%%%%%%%%%%%%%%%%%%%%%%%%%%%%%%%%%%%%%%%%
\partial_{z}\partial_{z}A_{t}&=\partial_{z}\left(\partial_{\theta}A_{\theta}+\frac{A_{\theta}}{\mathrm{tan}\theta}+\frac{\partial_{\phi}A_{\phi}}{\mathrm{sin}^{2}\theta}\right)-2\mathrm{Im}(\psi^{*}\partial_z\psi)\label{constraint_eq}\\
%%%%%%%%%%%%%%%%%%%%%%%%%%%%%%%%%%%%%%%%%%%%%%%%%%%%%%%%%%%%%%%%%%%%%%%%%%%%%%%%%%%
	\partial_{t}\partial_{z}A_{t}&=\frac{f \partial_{z}\partial_{\phi}A_{\phi}}{\mathrm{sin}^2\theta}+f\partial_{z}\partial_{\theta}A_{\theta}-\frac{\partial_{t}A_{\theta}-\partial_{\theta}A_{t}-f\partial_{z}A_{\theta}}{\mathrm{tan}\theta}-2A_{t}|\psi|^2+\partial^{2}_{\theta}A_t\notag\\ 
	&+i f (\psi^{*}\partial_{z}\psi-\psi\partial_{z}\psi^{*})+2\mathrm{Im}(\psi^{*}\partial_{t}\psi)+\frac{\partial_{\phi}(\partial_{\phi}A_t-\partial_{t}A_{\phi})}{\mathrm{sin}^2\theta}-\partial_{t}\partial_{\theta}A_{\theta}\\
%%%%%%%%%%%%%%%%%%%%%%%%%%%%%%%%%%%%%%%%%%%%%%%%%%%%%%%%%%%%%%%%%%%%%%%%%%%%%%%%%%%%%%%%%%%%%%%%%%%%%
	2\partial_{t}\partial_zA_{\theta}&=\partial_{z}(f\partial_{z}A_\theta+\partial_{\theta}A_{t})-2A_{\theta}|\psi|^2+2\mathrm{Im}(\psi^{*}\partial_{\theta}\psi)+\partial_{\phi}\frac{\partial_{\phi}A_{\theta}-\partial_{\theta}A_{\phi}}{\mathrm{sin}^2\theta}\label{theta}\\
%%%%%%%%%%%%%%%%%%%%%%%%%%%%%%%%%%%%%%%%%%%%%%%%%%%%%%%%%%%%%%%%%%%%%%%%%%%%%%%%%%%%%%%%%%%%%%%%%%%%
		2\partial_{t}\partial_zA_{\phi}&=\partial_{z}(f\partial_{z}A_\phi+\partial_{\phi}A_{t})-2A_{\phi}|\psi|^2+2\mathrm{Im}(\psi^{*}\partial_{\phi}\psi)+\frac{\partial_{\phi}A_{\theta}-\partial_{\theta}A_{\phi}}{\mathrm{tan}\theta}+\partial_{\theta}(\partial_{\theta}A_{\phi}-\partial_{\phi}A_{\theta})\label{phi}
	\end{align}

\begin{figure}
\centering
		\includegraphics[width=0.9\linewidth]{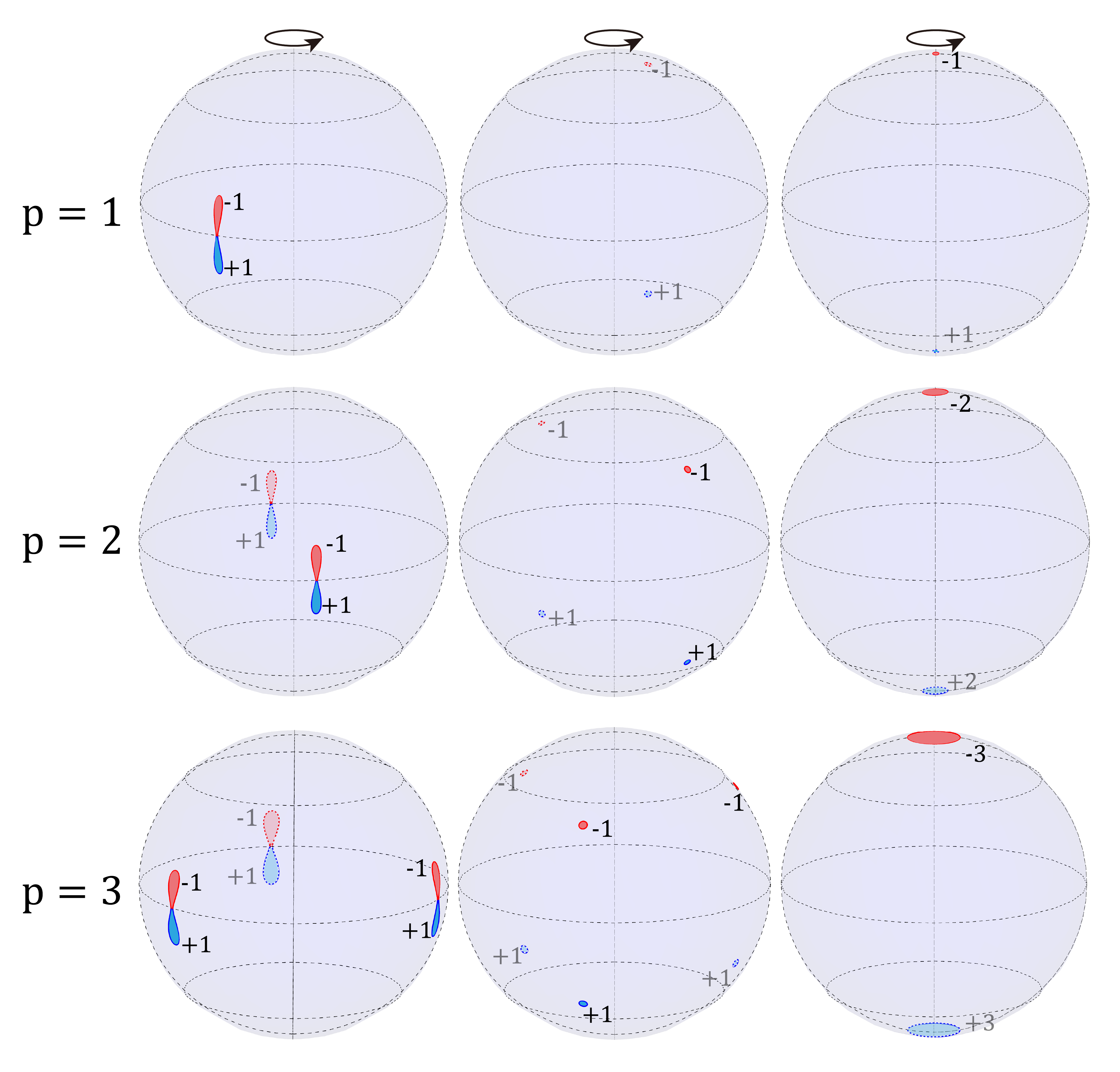}
		\caption{Dynamical evolution triggered by the dominant unstable even modes in the  $p=1$, $2$, and $3$ channels on the top of the superfluid configurations under $\Omega=3.8,5.2,6.6$, individually.} % (a) The evolution of  zones under positive angular velocity.  (b) The evolution of the condensate depletion zones under negative angular velocity.}
		\label{evolve1}
	\end{figure}

The initial data for the nonlinear evolution can be prepared by adding a perturbation to the static scalar field as follows
\begin{align}
\psi=\psi_{s}(z,\theta)+q_1(z,\theta)e^{i p\phi}+ q_2^{*}(z,\theta)e^{-i p\phi}
\end{align}
where $\psi_{s}(z,\theta)$ denotes the static superfluid configuration, and $q_1, q_2$ are the eigenvector associated with the QNMs. In addition, the initial $A_\theta$ and $A_\phi$ are given  by the static data, while the initial $A_t$ is determined by the constraint equation eq. (\ref{constraint_eq}).
We evolve the whole dynamics by eq. (\ref{Phi}), eq. (\ref{theta}), and eq. (\ref{phi}).

It is noteworthy that the above initial data from the perturbation by the single even and odd modes satisfy
\begin{align}
&\psi(\theta,\phi)=\psi(\pi-\theta,\phi+2\pi/p),\quad A_t(\theta,\phi)=A_t(\pi-\theta,\phi+2\pi/p),\notag\\
&A_\phi(\theta,\phi)=A_\phi(\pi-\theta,\phi+2\pi/p),\quad 
A_\theta(\theta,\phi)=-A_\theta(\pi-\theta,\phi+2\pi/p),
\end{align}
and 
\begin{align}
&\psi(\theta,\phi)=\psi(\pi-\theta,\phi+\pi/p),\quad A_t(\theta,\phi)=A_t(\pi-\theta,\phi+\pi/p),\notag\\
& A_\phi(\theta,\phi)=A_\phi(\pi-\theta,\phi+\pi/p),\quad A_\theta(\theta,\phi)=-A_\theta(\pi-\theta,\phi+\pi/p),
\end{align}
 respectively. Furthermore, the initial data are also invariant under a rotation by $2\pi/p$ in the azimuthal angle $\phi$. These symmetries are preserved throughout the nonlinear evolution and are clearly manifested in our numerical simulations.
  
 We first demonstrate the nonlinear evolution at positive $\Omega$ triggered by the dominant unstable even modes in the $p=1$, $2$, and $3$ channels, by perturbing the static superfluid configurations shown in figure~\ref{static}(b), (c), and (d), individually. The condensate initially exhibits distortions in the equatorial region, where the superfluid velocity reaches its maximum and vortex–antivortex pairs subsequently nucleate there due to the Landau instability. In figure~\ref{evolve1}, we show the condensate depletion zones surrounding the vortex (blue) and antivortex (red), defined as the regions enclosed by the 2\% contour of $|\langle O\rangle_{\text{max}}|$. We find that the mode number $p$ corresponds to the number of vortex–antivortex pairs nucleated in the equatorial region. Each vortex and antivortex subsequently separate, drift toward the poles, and spiral in the direction of the external rotation, eventually relaxing to a stationary vortex-antivortex configuration on the sphere. The corresponding order parameter is given by
 \begin{align}
 	\left\langle O\right\rangle=\left|\left\langle O\right\rangle\right| e^{i w \phi},
 \end{align}
where $w$ denotes winding number of the vortex located at the north pole, while the corresponding antivortex at the south pole carries winding number $-w$. For positive $\Omega$, the dominant unstable even mode in the $p$ channel leads to a final winding number $w=-p$, as shown in figure~\ref{evolve1}. As a result, we uncover the generic dynamical processes associated with the blue, red, and green regions for positive $\Omega$ in figure~\ref{QNM_N90}(d). These dynamical behaviors remain robust even when the initial perturbation is taken to be a random superposition of different modes, since the corresponding  dominant unstable even mode in each region eventually outgrows all other modes and dictates the late-time evolution.

 \begin{figure}
\centering
		\includegraphics[width=0.9\linewidth]{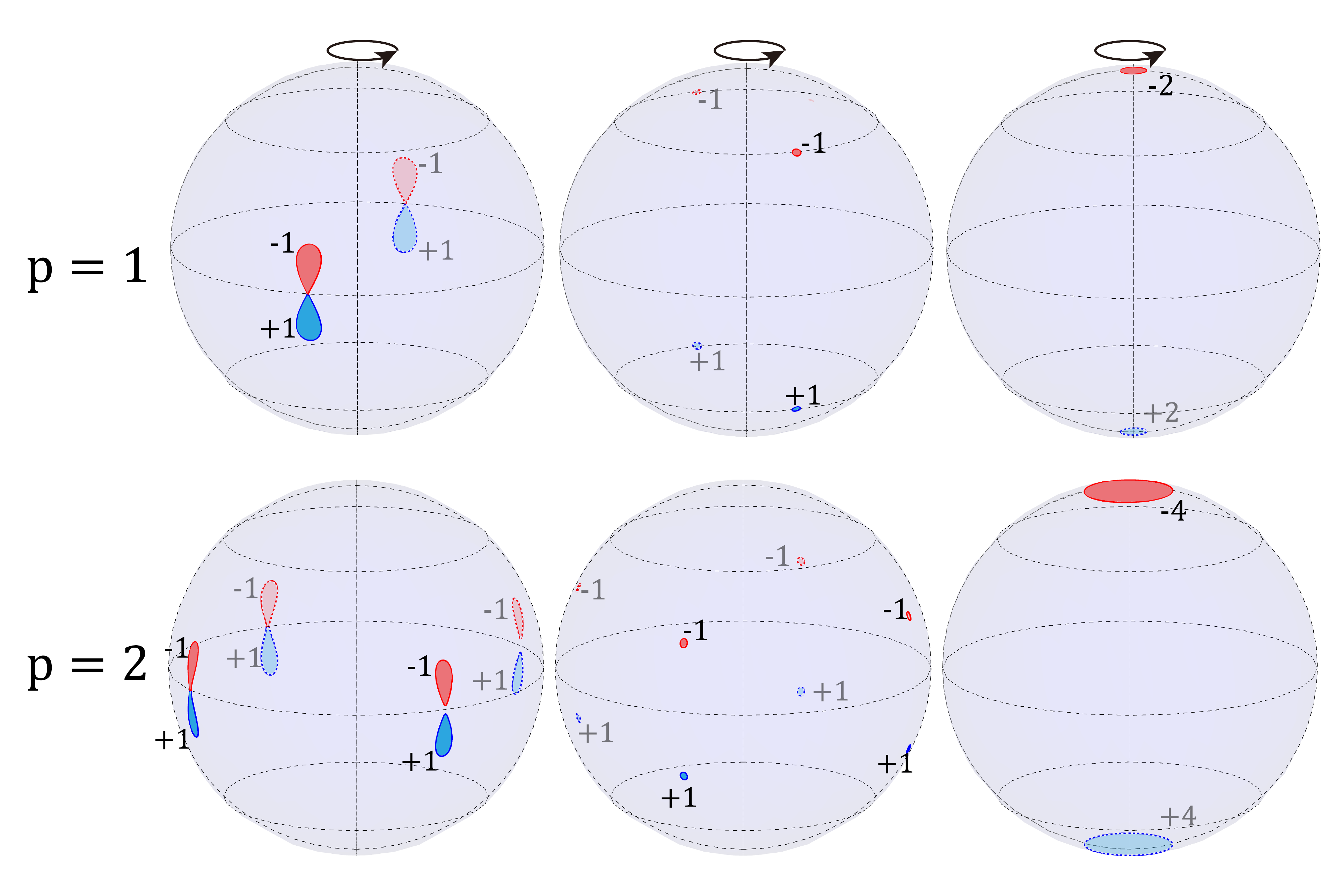}
		\caption{ Dynamical evolution triggered by unstable odd modes in the $p=1$ and $p=2$ channels on the top of the superfluid configurations under $\Omega=6.6$.} %(a) The evolution of the condensate depletion zones under positive $\Omega$.  (b) The evolution of the condensate depletion zones under negative $\Omega$.
		\label{evolve2}
	\end{figure} 
Next, we present the nonlinear evolution induced by the odd modes in the $p=1$ and $p=2$ channels by perturbing the static superfluid configurations shown in figure~\ref{static}(c) and (d), respectively. As shown in figure~\ref{evolve2}, $2p$ vortex--antivortex pairs are nucleated in the equatorial region, in contrast to the $p$ pairs produced by even perturbations. This behavior is a consequence of the symmetry property of the odd modes described in equation~(\ref{symmetry_for_odd_mode}). The system eventually relaxes into a stationary vortex--antivortex pair with winding number $w=-2p$.

In addition, we consider the case of negative $\Omega$. The dynamics in each $p$ channel likewise induce the nucleation of vortex--antivortex pairs in the equatorial region. However, these pairs carry topological charges opposite to those generated at the same $p$ for positive $\Omega$. This behavior is consistent with the analysis of Landau instabilities and soliton formation presented in \cite{g8sg-prdf}. After the vortices and antivortices separate and subsequently drift toward the poles, the system eventually relaxes into a stationary vortex-antivortex configuration with a winding number opposite to that obtained for positive $\Omega$.

 % newly emerged vortex–antivortex pair near the equator would reduce the local superfluid velocity between them. 

\section{Conclusion and discussion}\label{conclusion}
\begin{figure}
\centering
		\includegraphics[width=0.6\linewidth]{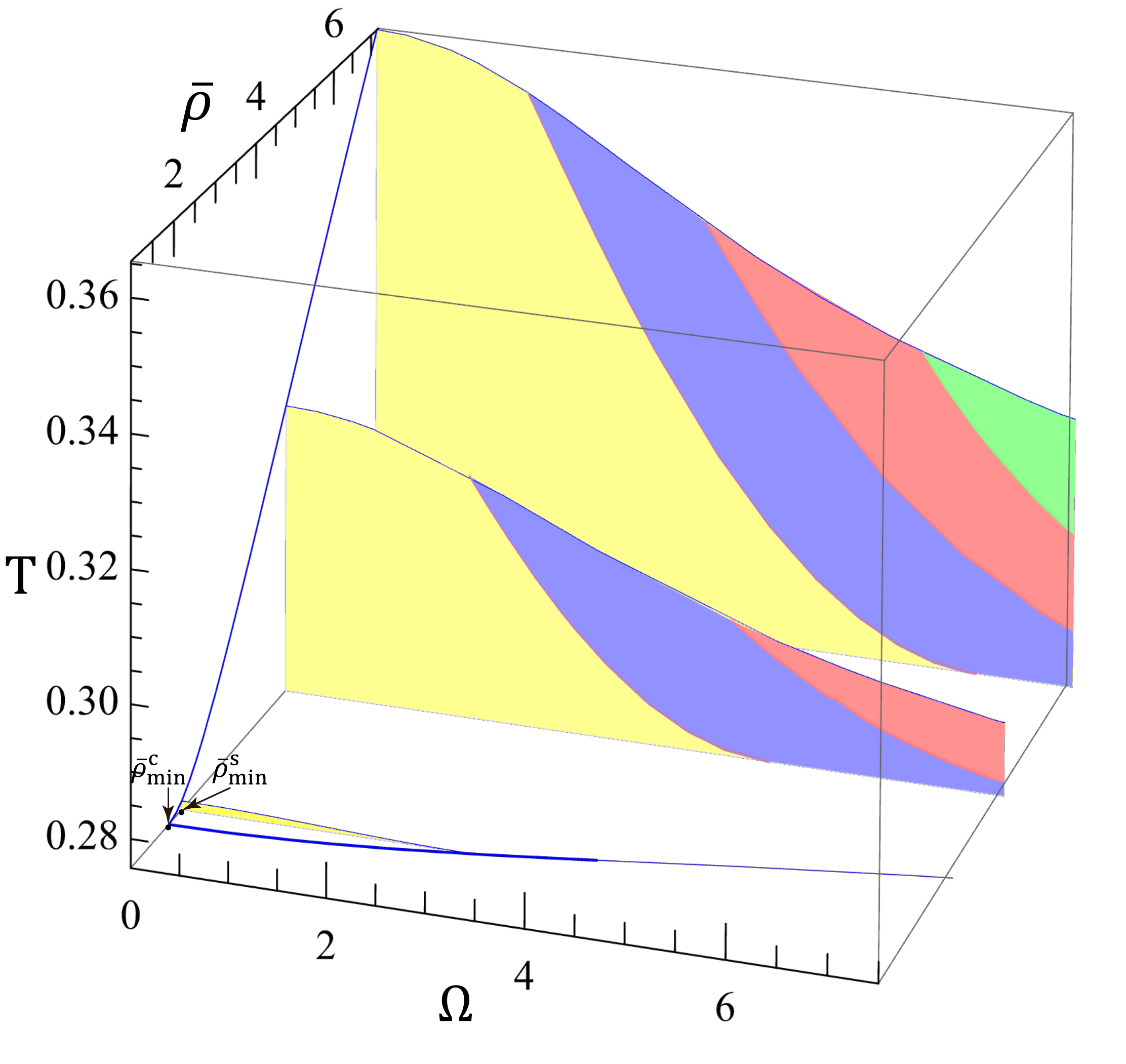}
		\caption{Dynamical phase diagram in the 3-dimensional phase space $(\Omega,\bar\rho,T)$ for the superfluid on the sphere. Critical average number density for the existence of superfluid solutions is found to be $\bar\rho^{c}_{\mathrm{min}}=1.891$ and the minimal number density that supports stable rotating shell superfluid is $\bar\rho^{s}_{\mathrm{min}}=2.1998$. The dynamical processes for the blue, red and green regions are controlled by the dominant unstable even modes in $p=1$, $2$, and $3$ channels, individually.}
		\label{3Dphase}
	\end{figure}

This work extends the study of uniform holographic superfluids on the sphere to inhomogeneous configurations under external rotation. Our analysis of the critical temperature reveals that it is universally enhanced on the spherical geometry compared with the planar case, regardless of the presence of external rotation. We further analyze the linear instability of the superfluid at fixed average number density $\bar\rho=22.5/\pi$ by computing QNMs. The unstable modes can be classified into even and odd sectors, with the even modes dominating and determining the dynamical phase boundaries in the $T$–$\Omega$ plane, as shown in figure \ref{QNM_N90}(d). Generally, by varying the average number density, we construct the dynamical phase diagram in the $(\Omega,\mu,T)$ parameter space, thereby revealing the complete structure of the dynamical phases, as shown in figure~\ref{3Dphase}. We find that the critical average number density for the existence of superfluid solutions is $\bar\rho^{c}_{\mathrm{min}}=1.891$, below which no superfluid phase exists. Moreover, the minimal average number density that supports stable rotating shell-shaped superfluid is $\bar\rho^{s}_{\mathrm{min}}=2.1998$. For $\bar\rho>\bar\rho^{s}_{\mathrm{min}}$, the superfluid remains stable in the small angular velocities but becomes dynamically unstable when the angular velocity exceeds a critical value.

We further perform fully nonlinear numerical simulations to investigate the ensuing dynamical evolution induced by different $p$-modes. The results show that vortex-antivortex pairs are nucleated at the equator, subsequently separate and drift toward the poles, and eventually accumulate there. The winding number ($w$) of the final vortex at the north pole depends on both the direction of rotation and the perturbation mode: for positive angular velocity, $w=-p$ for even modes and $w=-2p$ for odd modes, whereas the corresponding winding numbers reverse sign for negative angular velocity.

The phase transition phenomena and dynamical behaviors of the superfluid revealed in this work remain to be experimentally verified. Our results provide a foundation for exploring other inhomogeneous structures under finite rotation, such as solitons, vortices, skyrmions, and turbulence, within the framework of the AdS/CFT correspondence. Furthermore, it is worth noting an alternative holographic route to constructing a superfluid on a spherical geometry $S^2$. Instead of working directly in a spherical Schwarzschild-$\text{AdS}_4$ spacetime, one could construct a 3D shell-shaped condensate of finite thickness $\delta$ by applying a spherical confining potential on the boundary of $\text{AdS}_5$, and subsequently taking the dimensional reduction limit $\delta \to 0$. Although this embedding scheme is notoriously difficult to implement by numerics, it is particularly appealing from an experimental perspective, as it closely mirrors the physics of trapped shell-shaped BECs and provides a natural arena to explore the effect induced by the thickness $\delta$.  We leave this promising direction for future work.

\begin{acknowledgments}
This work is partially supported by the Zhiyuan Science
Foundation of BIPT with Grant No. 2026209 and the National Key Research and Development Program of China
with Grant No. 2021YFC2203001 as well as the National
Natural Science Foundation of China with Grant Nos.
12035016, 12275350, 12375048, 12375058, 12361141825,
12447182, and 12575047.

\end{acknowledgments}

\appendix

\section{Bulk profiles for the scalar and Maxwell fields}
\begin{figure}[h]
    \centering
    \includegraphics[width=\linewidth]{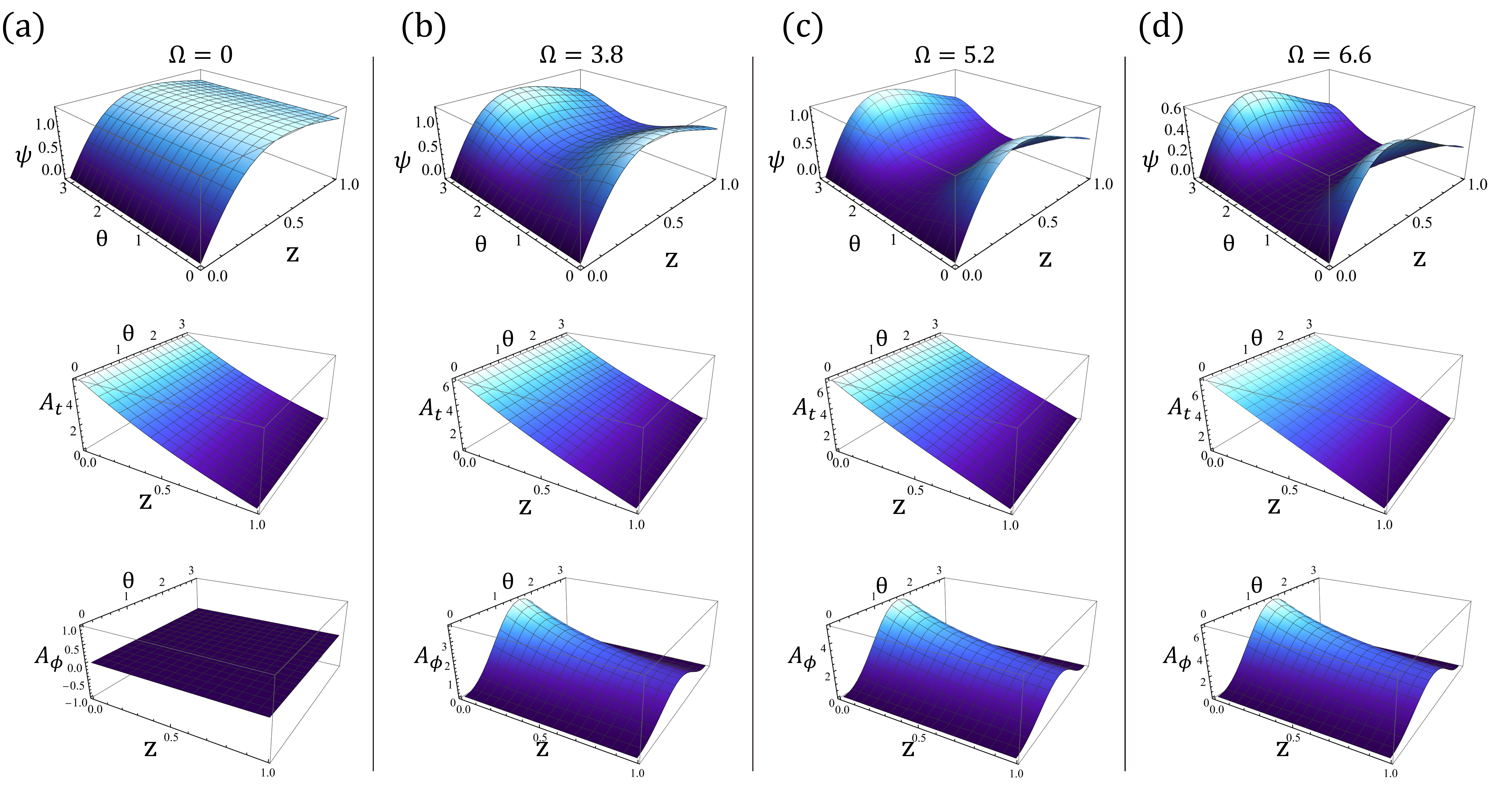}
    \caption{(a)-(d) show the bulk profiles of the scalar and Maxwell fields corresponding to the boundary superfluid for different angular velocities with $\Omega=0,3.8,5.2$ and $6.6$, individually.}
    \label{bulk_profile}
\end{figure}

In this appendix, we show the bulk profiles of the scalar and Maxwell fields in figure. \ref{bulk_profile}, which correspond to the different boundary superfluid configurations presented in figure.~\ref{static}.

\begin{figure}[h]
    \centering
    \includegraphics[width=\linewidth]{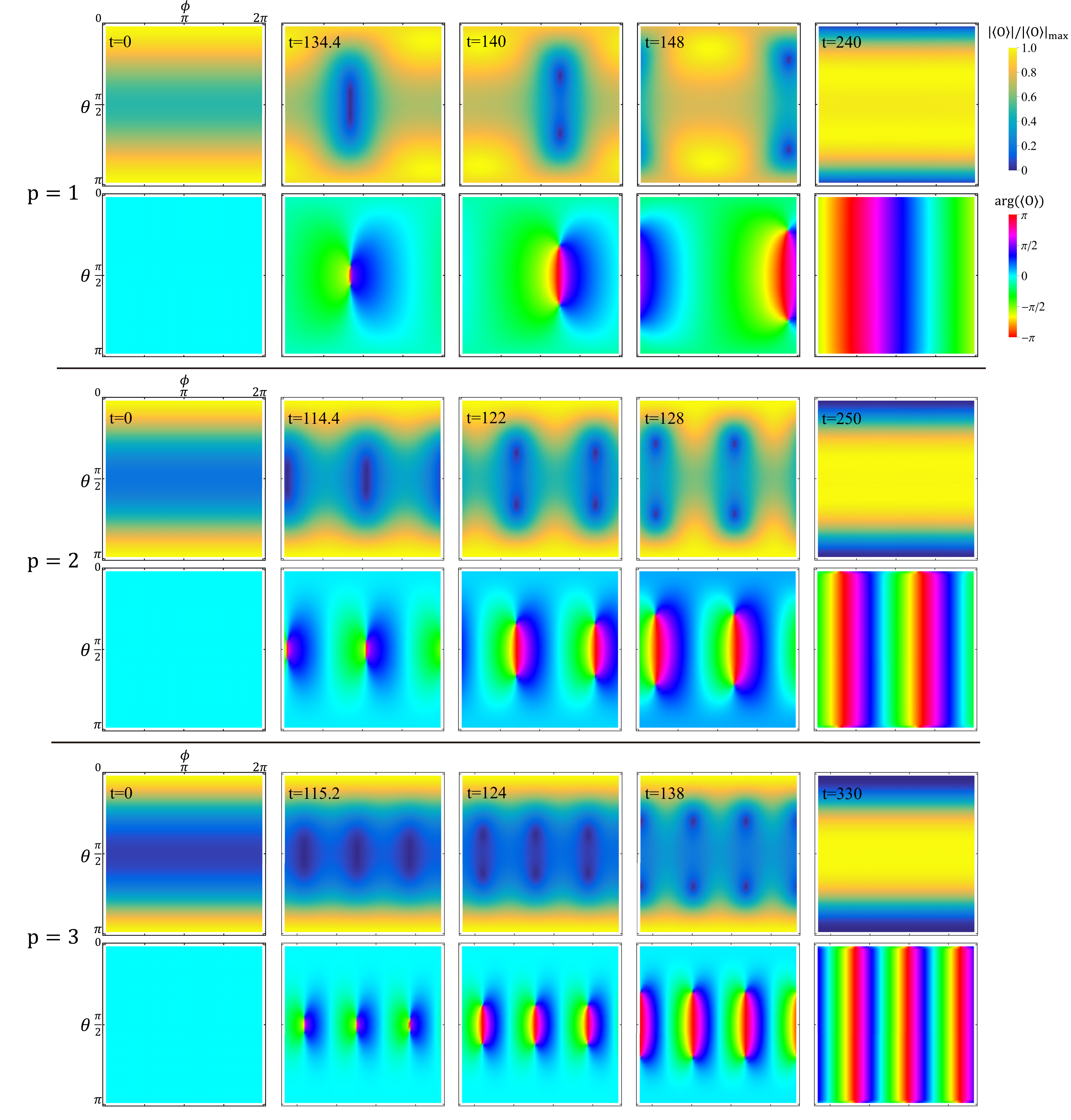}
    \caption{Full order parameter during the dynamical evolution triggered by the dominant unstable even modes in the  $p=1$, $2$, and $3$ channels on the top of the superfluid configurations under $\Omega=3.8,5.2,6.6$, individually.}
    \label{fulltime1}
\end{figure}

\section{Full Order parameter on the sphere during the non-linear time evolution}
\begin{figure}[h]
    \centering
    \includegraphics[width=\linewidth]{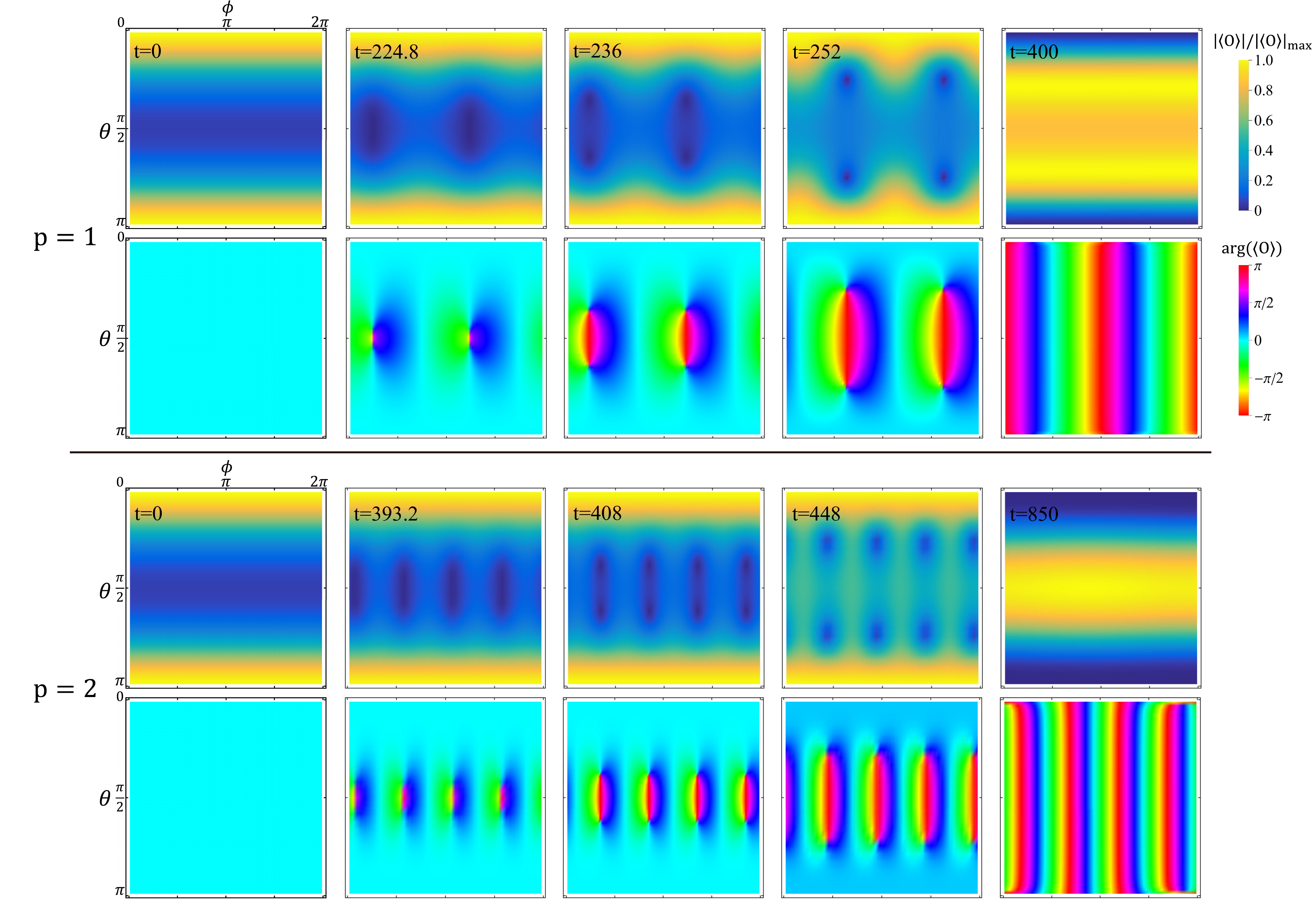}
    \caption{Full order parameter during the dynamical evolution triggered by unstable odd modes in the  $p=1$ and $p=2$ channels on the top of the superfluid configurations under $\Omega=6.6$.}
    \label{fulltime2}
\end{figure}

In this appendix, figures~\ref{fulltime1} and~\ref{fulltime2} present the full order-parameter configurations, showing both the normalized amplitude $|\langle O\rangle|/|\langle O\rangle|_{\mathrm{max}}$ and the phase $\arg(\langle O\rangle)$ over the entire spherical surface. These results complement the corresponding snapshots shown in figures~\ref{evolve1} and~\ref{evolve2}, respectively. To visualize the complete spatial distribution without self-occlusion, the configurations are represented in the $(\theta,\phi)$ plane.

\section{Discretization scheme in numerical time-dependent evolution}
In this appendix, we outline the spatial discretization scheme used in our numerical time-dependent evolutions.

For full time-dependent non-linear evolutions, we employ a Fourier spectral discretization in the azimuthal direction $\phi \in [0, 2\pi)$. In the $\theta$ direction, however, static polar boundary conditions cannot be imposed  in advance as was done for the static configuration, because the fields dynamically evolve over the entire surface. To address this, we extend the polar domain from $\theta \in [0, \pi]$ to $\theta \in (-\pi, \pi]$ and discretize it using a uniform Fourier grid with a half-grid shift:
\begin{align}
\theta_j = -\pi + \frac{(2j + 1)\pi}{N_\theta}, \qquad j = 0, 1, \dots, N_\theta - 1.
\end{align}
Consequently, no grid nodes coincide directly with the poles ($\theta = 0$ or $\theta = \pm\pi$), thereby avoiding the coordinate singularity ($\sin\theta = 0$) in the governing equations. Extending the domain to $\theta \in (-\pi, \pi]$ naturally enforces the double-cover parity symmetry on $S^2$:
\begin{align}
& \psi(\theta, \phi) = \psi(-\theta, \phi + \pi),
      \quad
   A_t(\theta, \phi) = A_t(-\theta, \phi + \pi) \notag\\
    \quad
 &   A_\phi(\theta, \phi) = A_{\phi}(-\theta, \phi + \pi),
    \quad
    A_\theta(\theta, \phi) = -A_{\theta}(-\theta, \phi + \pi).
\end{align}
This symmetry property allows us to reduce the size of the differentiation matrix in the $\theta$ direction from $N_\theta \times N_\theta$ to $(N_\theta/2) \times N_\theta$.

    \bibliographystyle{JHEP}
    \bibliography{Refs}

@article{allen1938flow,
  title={Flow of liquid helium II},
  author={Allen, John F and Misener, AD},
  journal={Nature},
  volume={141},
  number={3558},
  pages={75--75},
  year={1938},
  publisher={Nature Publishing Group UK London}
}

@article{kapitza1938viscosity,
  title={Viscosity of liquid helium below the $\lambda$-point},
  author={Kapitza, Pyotr},
  journal={Nature},
  volume={141},
  number={3558},
  pages={74--74},
  year={1938},
  publisher={Nature Publishing Group UK London}
}

@article{anderson1995observation,
  title={Observation of Bose-Einstein condensation in a dilute atomic vapor},
  author={Anderson, Mike H and Ensher, Jason R and Matthews, Michael R and Wieman, Carl E and Cornell, Eric A},
  journal={Science},
  volume={269},
  number={5221},
  pages={198--201},
  year={1995},
  publisher={American Association for the Advancement of Science}
}

@article{davis1995bose,
  title = {Bose-Einstein Condensation in a Gas of Sodium Atoms},
  author = {Davis, K. B. and Mewes, M. -O. and Andrews, M. R. and van Druten, N. J. and Durfee, D. S. and Kurn, D. M. and Ketterle, W.},
  journal = {Phys. Rev. Lett.},
  volume = {75},
  issue = {22},
  pages = {3969--3973},
  numpages = {0},
  year = {1995},
  month = {Nov},
  publisher = {American Physical Society}
}

@article{raman1999evidence,
  title={Evidence for a critical velocity in a Bose-Einstein condensed gas},
  author={Raman, C and K{\"o}hl, M and Onofrio, Roberto and Durfee, DS and Kuklewicz, CE and Hadzibabic, Z and Ketterle, W},
  journal={Phys. Rev. Lett.},
  volume={83},
  number={13},
  pages={2502},
  year={1999},
  publisher={American Physical Society}
}

@article{ramanathan2011superflow,
  title={Superflow in a toroidal Bose-Einstein condensate: An atom circuit with a tunable weak link},
  author={Ramanathan, Anand and Wright, KC and Muniz, S{\'e}rgio Ricardo and Zelan, Martin and Hill III, WT and Lobb, CJ and Helmerson, Kristian and Phillips, WD and Campbell, GK},
  journal={Phys. Rev. Lett.},
  volume={106},
  number={13},
  pages={130401},
  year={2011},
  publisher={APS}
}

@article{desbuquois2012superfluid,
  title={Superfluid behaviour of a two-dimensional Bose gas},
  author={Desbuquois, R{\'e}mi and Chomaz, Lauriane and Yefsah, Tarik and L{\'e}onard, Julian and Beugnon, J{\'e}r{\^o}me and Weitenberg, Christof and Dalibard, Jean},
  journal={Nat. Phys.},
  volume={8},
  number={9},
  pages={645--648},
  year={2012},
  publisher={Nature Publishing Group UK London}
}

@article{kwon2015periodic,
  title = {Periodic shedding of vortex dipoles from a moving penetrable obstacle in a Bose-Einstein condensate},
  author = {Kwon, Woo Jin and Seo, Sang Won and Shin, Y.},
  journal = {Phys. Rev. A},
  volume = {92},
  issue = {3},
  pages = {033613},
  numpages = {6},
  year = {2015},
  month = {Sep},
  publisher = {American Physical Society}
}

@article{kwon2016observation,
  title = {Observation of von K\'arm\'an Vortex Street in an Atomic Superfluid Gas},
  author = {Kwon, Woo Jin and Kim, Joon Hyun and Seo, Sang Won and Shin, Y.},
  journal = {Phys. Rev. Lett.},
  volume = {117},
  issue = {24},
  pages = {245301},
  numpages = {5},
  year = {2016},
  month = {Dec},
  publisher = {American Physical Society}
}

@article{madison2000vortex,
  title={Vortex formation in a stirred Bose-Einstein condensate},
  author={Madison, Kirk W and Chevy, Fr{\'e}d{\'e}ric and Wohlleben, Wendel and Dalibard, Jean},
  journal={Phys. Rev. Lett.},
  volume={84},
  number={5},
  pages={806},
  year={2000},
  publisher={APS}
}

@article{abo2001observation,
  title={Observation of vortex lattices in Bose-Einstein condensates},
  author={Abo-Shaeer, Jamil R and Raman, Chandra and Vogels, Johnny M and Ketterle, Wolfgang},
  journal={Science},
  volume={292},
  number={5516},
  pages={476--479},
  year={2001},
  publisher={American Association for the Advancement of Science}
}

@article{henn2009observation,
  title={Observation of vortex formation in an oscillating trapped Bose-Einstein condensate},
  author={Henn, EAL and Seman, JA and Ramos, ERF and Caracanhas, M and Castilho, P and Ol{\'\i}mpio, EP and Roati, G and Magalh{\~a}es, Daniel Varela and Magalh{\~a}es, K{\'\i}lvia Mayre Farias and Bagnato, Vanderlei Salvador},
  journal={Phys. Rev. A},
  volume={79},
  number={4},
  pages={043618},
  year={2009},
  publisher={APS}
}

@article{neely2010observation,
  title={Observation of vortex dipoles in an oblate Bose-Einstein condensate},
  author={Neely, Tyler W and Samson, Edward Carlo and Bradley, Ashton S and Davis, Matthew J and Anderson, Brian P},
  journal={Phys. Rev. Lett.},
  volume={104},
  number={16},
  pages={160401},
  year={2010},
  publisher={APS}
}

@article{miller2007critical,
  title={Critical velocity for superfluid flow across the BEC-BCS crossover},
  author={Miller, DE and Chin, JK and Stan, CA and Liu, Y and Setiawan, W and Sanner, C and Ketterle, W},
  journal={Phys. Rev. Lett.},
  volume={99},
  number={7},
  pages={070402},
  year={2007},
  publisher={APS}
}

@article{wright2013threshold,
  title={Threshold for creating excitations in a stirred superfluid ring},
  author={Wright, KC and Blakestad, RB and Lobb, CJ and Phillips, WD and Campbell, GK},
  journal={Phys. Rev. A},
  volume={88},
  number={6},
  pages={063633},
  year={2013},
  publisher={APS}
}

@article{carollo2022observation,
  title={Observation of ultracold atomic bubbles in orbital microgravity},
  author={Carollo, Ryan A and Aveline, David C and Rhyno, Brendan and Vishveshwara, Smitha and Lannert, Courtney and Murphree, Joseph D and Elliott, Ethan R and Williams, Jason R and Thompson, Robert J and Lundblad, Nathan},
  journal={Nature},
  volume={606},
  number={7913},
  pages={281--286},
  year={2022},
  publisher={Nature Publishing Group UK London}
}

@article{jia2022expansion,
  title={Expansion dynamics of a shell-shaped Bose-Einstein condensate},
  author={Jia, Fan and Huang, Zerong and Qiu, Liyuan and Zhou, Rongzi and Yan, Yangqian and Wang, Dajun},
  journal={Phys. Rev. Lett.},
  volume={129},
  number={24},
  pages={243402},
  year={2022},
  publisher={APS}
}

@article{lundblad2023perspective,
  title={Perspective on quantum bubbles in microgravity},
  author={Lundblad, Nathan and Aveline, David C and Bala{\v{z}}, Antun and Bentine, Elliot and Bigelow, Nicholas P and Boegel, Patrick and Efremov, Maxim A and Gaaloul, Naceur and Meister, Matthias and Olshanii, Maxim and others},
  journal={Quantum Sci. Technol.},
  volume={8},
  number={2},
  pages={024003},
  year={2023},
  publisher={IOP Publishing}
}

@article{lundblad2019shell,
  title={Shell potentials for microgravity Bose--Einstein condensates},
  author={Lundblad, Nathan and Carollo, RA and Lannert, Courtney and Gold, MJ and Jiang, X and Paseltiner, D and Sergay, N and Aveline, DC},
  journal={npj Microgravity},
  volume={5},
  number={1},
  pages={30},
  year={2019},
  publisher={Nature Publishing Group UK London}
}

@article{sun2018static,
  title={Static and dynamic properties of shell-shaped condensates},
  author={Sun, Kuei and Padavi{\'c}, Karmela and Yang, Frances and Vishveshwara, Smitha and Lannert, Courtney},
  journal={Phys. Rev. A},
  volume={98},
  number={1},
  pages={013609},
  year={2018},
  publisher={APS}
}

@article{rhyno2021thermodynamics,
  title={Thermodynamics in expanding shell-shaped Bose-Einstein condensates},
  author={Rhyno, Brendan and Lundblad, Nathan and Aveline, David C and Lannert, Courtney and Vishveshwara, Smitha},
  journal={Phys. Rev. A},
  volume={104},
  number={6},
  pages={063310},
  year={2021},
  publisher={APS}
}

@article{tononi2019bose,
  title={Bose-Einstein condensation on the surface of a sphere},
  author={Tononi, A and Salasnich, L},
  journal={Phys. Rev. Lett.},
  volume={123},
  number={16},
  pages={160403},
  year={2019},
  publisher={APS}
}

@article{tononi2020quantum,
  title={Quantum bubbles in microgravity},
  author={Tononi, A and Cinti, F and Salasnich, L},
  journal={Phys. Rev. Lett.},
  volume={125},
  number={1},
  pages={010402},
  year={2020},
  publisher={APS}
}

@article{moller2020bose,
  title={Bose--Einstein condensation on curved manifolds},
  author={M{\'o}ller, Nat{\'a}lia S and dos Santos, F Ednilson A and Bagnato, Vanderlei S and Pelster, Axel},
  journal={New J. Phys.},
  volume={22},
  number={6},
  pages={063059},
  year={2020},
  publisher={IOP Publishing}
}

@article{tononi2022scattering,
  title={Scattering theory and equation of state of a spherical two-dimensional Bose gas},
  author={Tononi, A},
  journal={Physical Review A},
  volume={105},
  number={2},
  pages={023324},
  year={2022},
  publisher={APS}
}

@article{lannert2007dynamics,
  title={Dynamics of condensate shells: Collective modes and expansion},
  author={Lannert, Courtney and Wei, T-C and Vishveshwara, Smitha},
  journal={Physical Review A—Atomic, Molecular, and Optical Physics},
  volume={75},
  number={1},
  pages={013611},
  year={2007},
  publisher={APS}
}

@article{padavic2017physics,
  title={Physics of hollow Bose-Einstein condensates},
  author={Padavi{\'c}, Karmela and Sun, Kuei and Lannert, Courtney and Vishveshwara, Smitha},
  journal={Europhysics Letters},
  volume={120},
  number={2},
  pages={20004},
  year={2017},
  publisher={EDP Sciences, IOP Publishing and Societ{\`a} Italiana di Fisica}
}

@article{boegel2023controlled,
  title={Controlled expansion of shell-shaped Bose--Einstein condensates},
  author={Boegel, Patrick and Wolf, Alexander and Meister, Matthias and Efremov, Maxim A},
  journal={Quantum Science and Technology},
  volume={8},
  number={3},
  pages={034001},
  year={2023},
  publisher={IOP Publishing}
}

@article{greiner2002quantum,
  title={Quantum phase transition from a superfluid to a Mott insulator in a gas of ultracold atoms},
  author={Greiner, Markus and Mandel, Olaf and Esslinger, Tilman and H{\"a}nsch, Theodor W and Bloch, Immanuel},
  journal={nature},
  volume={415},
  number={6867},
  pages={39--44},
  year={2002},
  publisher={Nature Publishing Group UK London}
}

@article{seaman2007atomtronics,
  title={Atomtronics: Ultracold-atom analogs of electronic devices},
  author={Seaman, BT and Kr{\"a}mer, M and Anderson, DZ and Holland, MJ},
  journal={Physical Review A—Atomic, Molecular, and Optical Physics},
  volume={75},
  number={2},
  pages={023615},
  year={2007},
  publisher={APS}
}

@article{bloch2012quantum,
  title={Quantum simulations with ultracold quantum gases},
  author={Bloch, Immanuel and Dalibard, Jean and Nascimbene, Sylvain},
  journal={Nature Physics},
  volume={8},
  number={4},
  pages={267--276},
  year={2012},
  publisher={Nature Publishing Group}
}

@article{greiner2003emergence,
  title={Emergence of a molecular Bose--Einstein condensate from a Fermi gas},
  author={Greiner, Markus and Regal, Cindy A and Jin, Deborah S},
  journal={Nature},
  volume={426},
  number={6966},
  pages={537--540},
  year={2003},
  publisher={Nature Publishing Group UK London}
}

@article{zwierlein2005vortices,
  title={Vortices and superfluidity in a strongly interacting Fermi gas},
  author={Zwierlein, Martin W and Abo-Shaeer, Jamil R and Schirotzek, Andre and Schunck, Christian H and Ketterle, Wolfgang},
  journal={Nature},
  volume={435},
  number={7045},
  pages={1047--1051},
  year={2005},
  publisher={Nature Publishing Group UK London}
}

@article{giorgini2008theory,
  title={Theory of ultracold atomic Fermi gases},
  author={Giorgini, Stefano and Pitaevskii, Lev P and Stringari, Sandro},
  journal={Reviews of Modern Physics},
  volume={80},
  number={4},
  pages={1215--1274},
  year={2008},
  publisher={APS}
}

@article{steinhauer2016observation,
  title={Observation of quantum Hawking radiation and its entanglement in an analogue black hole},
  author={Steinhauer, Jeff},
  journal={Nature Physics},
  volume={12},
  number={10},
  pages={959--965},
  year={2016},
  publisher={Nature Publishing Group UK London}
}

@article{pines1985superfluidity,
  title={Superfluidity in neutron stars},
  author={Pines, David and Alpar, M Ali},
  journal={Nature},
  volume={316},
  number={6023},
  pages={27--32},
  year={1985},
  publisher={Nature Publishing Group UK London}
}

@article{almeida2023analogue,
  title={Analogue gravity and the Hawking effect: historical perspective and literature review},
  author={Almeida, Carla R and Jacquet, Maxime J},
  journal={The European Physical Journal H},
  volume={48},
  number={1},
  pages={15},
  year={2023},
  publisher={Springer}
}

@article{elliott2018nasa,
  title={NASA’s Cold Atom Lab (CAL): system development and ground test status},
  author={Elliott, Ethan R and Krutzik, Markus C and Williams, Jason R and Thompson, Robert J and Aveline, David C},
  journal={npj Microgravity},
  volume={4},
  number={1},
  pages={16},
  year={2018},
  publisher={Nature Publishing Group UK London}
}

@article{aveline2020observation,
  title={Observation of Bose--Einstein condensates in an Earth-orbiting research lab},
  author={Aveline, David C and Williams, Jason R and Elliott, Ethan R and Dutenhoffer, Chelsea and Kellogg, James R and Kohel, James M and Lay, Norman E and Oudrhiri, Kamal and Shotwell, Robert F and Yu, Nan and others},
  journal={Nature},
  volume={582},
  number={7811},
  pages={193--197},
  year={2020},
  publisher={Nature Publishing Group UK London}
}

@article{van2010bose,
  title={Bose-Einstein condensation in microgravity},
  author={van Zoest, Tim and Gaaloul, Naceur and Singh, Yeshpal and Ahlers, Holger and Herr, W and Seidel, ST and Ertmer, Wolfgang and Rasel, E and Eckart, Michael and Kajari, Endre and others},
  journal={Science},
  volume={328},
  number={5985},
  pages={1540--1543},
  year={2010},
  publisher={American Association for the Advancement of Science}
}

@article{muntinga2013interferometry,
  title={Interferometry with Bose-Einstein condensates in microgravity},
  author={M{\"u}ntinga, H and Ahlers, Hendrik and Krutzik, M and Wenzlawski, A and Arnold, Stefan and Becker, D and Bongs, K and Dittus, H and Duncker, H and Gaaloul, N and others},
  journal={Physical review letters},
  volume={110},
  number={9},
  pages={093602},
  year={2013},
  publisher={APS}
}

@article{vogt2020evaporative,
  title={Evaporative cooling from an optical dipole trap in microgravity},
  author={Vogt, Christian and Woltmann, Marian and Herrmann, Sven and L{\"a}mmerzahl, Claus and Albers, Henning and Schlippert, Dennis and Rasel, Ernst M and PRIMUS},
  journal={Physical Review A},
  volume={101},
  number={1},
  pages={013634},
  year={2020},
  publisher={APS}
}

@article{becker2018space,
  title={Space-borne Bose--Einstein condensation for precision interferometry},
  author={Becker, Dennis and Lachmann, Maike D and Seidel, Stephan T and Ahlers, Holger and Dinkelaker, Aline N and Grosse, Jens and Hellmig, Ortwin and M{\"u}ntinga, Hauke and Schkolnik, Vladimir and Wendrich, Thijs and others},
  journal={Nature},
  volume={562},
  number={7727},
  pages={391--395},
  year={2018},
  publisher={Nature Publishing Group UK London}
}

@article{condon2019all,
  title={All-optical Bose-Einstein condensates in microgravity},
  author={Condon, Gabriel and Rabault, Martin and Barrett, Brynle and Chichet, Laure and Arguel, Romain and Eneriz-Imaz, Hodei and Naik, Devang and Bertoldi, Andrea and Battelier, Baptiste and Bouyer, P and others},
  journal={Physical Review Letters},
  volume={123},
  number={24},
  pages={240402},
  year={2019},
  publisher={APS}
}

@article{kanai2021true,
  title={True mechanism of spontaneous order from turbulence in two-dimensional superfluid manifolds},
  author={Kanai, Toshiaki and Guo, Wei},
  journal={Phys. Rev. Lett.},
  volume={127},
  number={9},
  pages={095301},
  year={2021},
  publisher={APS}
}

@article{bereta2021superfluid,
  title={Superfluid vortex dynamics on a spherical film},
  author={Bereta, S{\'a}lvio J and Caracanhas, M{\^o}nica A and Fetter, Alexander L},
  journal={Phys. Rev. A},
  volume={103},
  number={5},
  pages={053306},
  year={2021},
  publisher={APS}
}

@article{padavic2020vortex,
  title={Vortex-antivortex physics in shell-shaped Bose-Einstein condensates},
  author={Padavi{\'c}, Karmela and Sun, Kuei and Lannert, Courtney and Vishveshwara, Smitha},
  journal={Phys. Rev. A},
  volume={102},
  number={4},
  pages={043305},
  year={2020},
  publisher={APS}
}

@article{white2024triangular,
  title={Triangular vortex lattices and giant vortices in rotating bubble Bose-Einstein condensates},
  author={White, Angela C},
  journal={Phys. Rev. A},
  volume={109},
  number={1},
  pages={013301},
  year={2024},
  publisher={APS}
}

@article{tononi2022topological,
  title={Topological superfluid transition in bubble-trapped condensates},
  author={Tononi, Andrea and Pelster, Axel and Salasnich, Luca},
  journal={Phys. Rev. Res.},
  volume={4},
  number={1},
  pages={013122},
  year={2022},
  publisher={APS}
}

@article{li2023equatorial,
  title={Equatorial waves in rotating bubble-trapped superfluids},
  author={Li, Guangyao and Efimkin, Dmitry K},
  journal={Phys. Rev. A},
  volume={107},
  number={2},
  pages={023319},
  year={2023},
  publisher={APS}
}

@article{saito2023rossby,
  title={Rossby--Haurwitz wave in a rotating bubble-shaped Bose--Einstein condensate},
  author={Saito, Hiroki and Hayashi, Masazumi},
  journal={Journal of the Physical Society of Japan},
  volume={92},
  number={4},
  pages={044003},
  year={2023},
  publisher={The Physical Society of Japan}
}

@article{ciardi2024supersolid,
  title={Supersolid phases of bosonic particles in a bubble trap},
  author={Ciardi, Matteo and Cinti, Fabio and Pellicane, Giuseppe and Prestipino, Santi},
  journal={Phys. Rev. Lett.},
  volume={132},
  number={2},
  pages={026001},
  year={2024},
  publisher={APS}
}

@article{gao2025holographic,
  title={Holographic homogeneous superfluid on the sphere},
  author={Gao, Meng and Ning, Zhuan and Tian, Yu and Zhang, Hongbao},
  journal={J. High Energy Phys.},
  volume={2025},
  number={2},
  pages={1--20},
  year={2025},
  publisher={Springer}
}

@article{PhysRevLett.101.031601,
  title = {Building a Holographic Superconductor},
  author = {Hartnoll, Sean A. and Herzog, Christopher P. and Horowitz, Gary T.},
  journal = {Phys. Rev. Lett.},
  volume = {101},
  issue = {3},
  pages = {031601},
  numpages = {4},
  year = {2008},
  month = {Jul},
  publisher = {American Physical Society}
}

@article{hartnoll2008holographic,
  title={Holographic superconductors},
  author={Hartnoll, Sean A and Herzog, Christopher P and Horowitz, Gary T},
  journal={J. High Energy Phys.},
  volume={2008},
  number={12},
  pages={015},
  year={2008},
  publisher={IOP Publishing}
}

@article{liu2019holographic,
  title={Holographic systems far from equilibrium: a review},
  author={Liu, Hong and Sonner, Julian},
  journal={Rep. Prog. Phys.},
  volume={83},
  number={1},
  pages={016001},
  year={2019},
  publisher={IOP Publishing}
}

@article{PhysRevD.79.066002,
  title = {Holographic model of superfluidity},
  author = {Herzog, C. P. and Kovtun, P. K. and Son, D. T.},
  journal = {Phys. Rev. D},
  volume = {79},
  issue = {6},
  pages = {066002},
  numpages = {11},
  year = {2009},
  month = {Mar},
  publisher = {American Physical Society}
}

@article{PhysRevD.78.065034,
  title = {Breaking an Abelian gauge symmetry near a black hole horizon},
  author = {Gubser, Steven S.},
  journal = {Phys. Rev. D},
  volume = {78},
  issue = {6},
  pages = {065034},
  numpages = {7},
  year = {2008},
  month = {Sep},
  publisher = {American Physical Society}
}

@article{hohenberg1977theory,
  title={Theory of dynamic critical phenomena},
  author={Hohenberg, Pierre C and Halperin, Bertrand I},
  journal={Rev. Mod. Phys.},
  volume={49},
  number={3},
  pages={435},
  year={1977},
  publisher={APS}
}

@article{keranen2010inhomogeneoussoliton,
  title={Inhomogeneous structures in holographic superfluids. I. Dark solitons},
  author={Ker{\"a}nen, Ville and Keski-Vakkuri, Esko and Nowling, Sean and Yogendran, KP},
  journal={Phys. Rev. D},
  volume={81},
  number={12},
  pages={126011},
  year={2010},
  publisher={APS}
}

@article{keranen2010inhomogeneous,
  title={Inhomogeneous structures in holographic superfluids. II. Vortices},
  author={Ker{\"a}nen, Ville and Keski-Vakkuri, Esko and Nowling, Sean and Yogendran, KP},
  journal={Phys. Rev. D},
  volume={81},
  number={12},
  pages={126012},
  year={2010},
  publisher={APS}
}

@article{domenech2010emergent,
  title={Emergent gauge fields in holographic superconductors},
  author={Domenech, Oriol and Montull, Marc and Pomarol, Alex and Salvio, Alberto and Silva, Pedro J},
  journal={J. High Energy Phys.},
  volume={2010},
  number={8},
  pages={1--20},
  year={2010},
  publisher={Springer}
}

@article{keranen2011solitons,
  title={Solitons as probes of the structure of holographic superfluids},
  author={Ker{\"a}nen, Ville and Keski-Vakkuri, Esko and Nowling, Sean and Yogendran, KP},
  journal={New J. Phys.},
  volume={13},
  number={6},
  pages={065003},
  year={2011},
  publisher={IOP Publishing}
}

@article{montull2012magnetic,
  title={Magnetic response in the holographic insulator/superconductor transition},
  author={Montull, Marc and Pujolas, Oriol and Salvio, Alberto and Silva, Pedro J},
  journal={J. High Energy Phys.},
  volume={2012},
  number={4},
  pages={1--34},
  year={2012},
  publisher={Springer}
}

@article{xia2019vortex,
  title={Vortex lattice in a rotating holographic superfluid},
  author={Xia, Chuan-Yin and Zeng, Hua-Bi and Zhang, Hai-Qing and Nie, Zhang-Yu and Tian, Yu and Li, Xin},
  journal={Phys. Rev. D},
  volume={100},
  number={6},
  pages={061901},
  year={2019},
  publisher={APS}
}

@article{lan2019attractive,
  title={Attractive interaction between vortex and anti-vortex in holographic superfluid},
  author={Lan, Shan-Quan and Li, Gu-Qiang and Mo, Jie-Xiong and Xu, Xiao-Bao},
  journal={J. High Energy Phys.},
  volume={2019},
  number={2},
  pages={1--15},
  year={2019},
  publisher={Springer}
}

@article{guo2020dynamical,
  title={Dynamical phase transition from nonequilibrium dynamics of dark solitons},
  author={Guo, Minyong and Keski-Vakkuri, Esko and Liu, Hong and Tian, Yu and Zhang, Hongbao},
  journal={Phys. Rev. Lett.},
  volume={124},
  number={3},
  pages={031601},
  year={2020},
  publisher={APS}
}

@article{li2020generation,
  title={Generation of vortices and stabilization of vortex lattices in holographic superfluids},
  author={Li, Xin and Tian, Yu and Zhang, Hongbao},
  journal={J. High Energy Phys.},
  volume={2020},
  number={2},
  pages={1--19},
  year={2020},
  publisher={Springer}
}

@article{wittmer2021vortex,
  title={Vortex motion quantifies strong dissipation in a holographic superfluid},
  author={Wittmer, Paul and Schmied, Christian-Marcel and Gasenzer, Thomas and Ewerz, Carlo},
  journal={Phys. Rev. Lett.},
  volume={127},
  number={10},
  pages={101601},
  year={2021},
  publisher={APS}
}

@article{ewerz2021dynamics,
  title={Dynamics of a vortex dipole in a holographic superfluid},
  author={Ewerz, Carlo and Samberg, Andreas and Wittmer, Paul},
  journal={J. High Energy Phys.},
  volume={2021},
  number={11},
  pages={1--46},
  year={2021},
  publisher={Springer}
}

@article{yan2023holographic,
  title={Holographic dissipation prefers the Landau over the Keldysh form},
  author={Yan, Yu-Kun and Lan, Shanquan and Tian, Yu and Yang, Peng and Yao, Shunhui and Zhang, Hongbao},
  journal={Phys. Rev. D},
  volume={107},
  number={12},
  pages={L121901},
  year={2023},
  publisher={APS}
}

@article{lan2023splitting,
  title={Splitting of doubly quantized vortices in holographic superfluid of finite temperature},
  author={Lan, Shanquan and Li, Xin and Mo, Jiexiong and Tian, Yu and Yan, Yu-Kun and Yang, Peng and Zhang, Hongbao},
  journal={J. High Energy Phys.},
  volume={2023},
  number={5},
  pages={1--17},
  year={2023},
  publisher={Springer}
}

@article{lan2023heating,
  title={Heating up quadruply quantized vortices: Splitting patterns and dynamical transitions},
  author={Lan, Shanquan and Li, Xin and Tian, Yu and Yang, Peng and Zhang, Hongbao},
  journal={Phys. Rev. Lett.},
  volume={131},
  number={22},
  pages={221602},
  year={2023},
  publisher={APS}
}

@article{yang2023motion,
  title={Motion of a superfluid vortex according to holographic quantum dissipation},
  author={Yang, Wei-Can and Xia, Chuan-Yin and Zeng, Hua-Bi and Tsubota, Makoto and Zaanen, Jan},
  journal={Phys. Rev. B},
  volume={107},
  number={14},
  pages={144511},
  year={2023},
  publisher={APS}
}

@article{su2023giant,
  title={Giant vortex in a fast rotating holographic superfluid},
  author={Su, Jia-Hao and Xia, Chuan-Yin and Yang, Wei-Can and Zeng, Hua-Bi},
  journal={Phys. Rev. D},
  volume={107},
  number={2},
  pages={026006},
  year={2023},
  publisher={APS}
}

@article{chesler2013holographic,
  title={Holographic vortex liquids and superfluid turbulence},
  author={Chesler, Paul M and Liu, Hong and Adams, Allan},
  journal={Science},
  volume={341},
  number={6144},
  pages={368--372},
  year={2013},
  publisher={American Association for the Advancement of Science}
}

@article{du2015holographic,
  title={Holographic thermal relaxation in superfluid turbulence},
  author={Du, Yiqiang and Niu, Chao and Tian, Yu and Zhang, Hongbao},
  journal={J. High Energy Phys.},
  volume={2015},
  number={12},
  pages={1--12},
  year={2015},
  publisher={Springer}
}

@article{lan2016towards,
  title={Towards quantum turbulence in finite temperature Bose-Einstein condensates},
  author={Lan, Shanquan and Tian, Yu and Zhang, Hongbao},
  journal={J. High Energy Phys.},
  volume={2016},
  number={7},
  pages={1--16},
  year={2016},
  publisher={Springer}
}

@article{ewerz2015non,
  title={Non-thermal fixed point in a holographic superfluid},
  author={Ewerz, Carlo and Gasenzer, Thomas and Karl, Markus and Samberg, Andreas},
  journal={J. High Energy Phys.},
  volume={2015},
  number={5},
  pages={1--36},
  year={2015},
  publisher={Springer}
}

@article{yang2024mechanism,
  title={Mechanism for cluster formation in a strongly interacting superfluid from gauge/gravity duality},
  author={Yang, Wei-Can and Xia, Chuan-Yin and Tian, Yu and Tsubota, Makoto and Zeng, Hua-Bi},
  journal={Phys. Rev. B},
  volume={110},
  number={13},
  pages={134510},
  year={2024},
  publisher={APS}
}

@article{zeng2025dissipation,
  title={Dissipation and decay of three-dimensional holographic quantum turbulence},
  author={Zeng, Hua-Bi and Xia, Chuan-Yin and Yang, Wei-Can and Tian, Yu and Tsubota, Makoto},
  journal={Phys. Rev. Lett.},
  volume={134},
  number={9},
  pages={091603},
  year={2025},
  publisher={APS}
}

@article{amado2014holographic,
  title={Holographic superfluids and the Landau criterion},
  author={Amado, Irene and Are{\'a}n, Daniel and Jim{\'e}nez-Alba, Amadeo and Landsteiner, Karl and Melgar, Luis and Landea, Ignacio Salazar},
  journal={J. High Energy Phys.},
  volume={2014},
  number={2},
  pages={1--24},
  year={2014},
  publisher={Springer}
}

@article{gouteraux2023critical,
  title={Critical superflows and thermodynamic instabilities in superfluids},
  author={Gout{\'e}raux, Blaise and Sottovia, Filippo and Mefford, Eric},
  journal={Phys. Rev. D},
  volume={108},
  number={8},
  pages={L081903},
  year={2023},
  publisher={APS}
}

@article{g8sg-prdf,
  title = {Landau instability and soliton formations},
  author = {Lan, Shanquan and Liu, Hong and Tian, Yu and Zhang, Hongbao},
  journal = {Phys. Rev. D},
  volume = {112},
  issue = {2},
  pages = {L021901},
  numpages = {8},
  year = {2025},
  month = {Jul},
  publisher = {American Physical Society}
}
\end{document}